\documentclass[preprint,aps,prd,superscriptaddress,amssymb,amsmath,nofootinbib]{revtex4-1}

\usepackage[linktocpage,pagebackref=true]{hyperref}

\usepackage{amsmath,setspace,amsfonts,latexsym}
\usepackage{amssymb}
\usepackage{color}
\usepackage{epsfig}
\usepackage{graphicx}
\usepackage{slashed}
\usepackage{caption}
\usepackage{subcaption}

\definecolor{White}{rgb}{1,1,1}
\definecolor{Red}{rgb}{1,0.1,0}
\definecolor{LightYellow}{rgb}{1,1,.875}
\definecolor{SteelBlue}{rgb}{.273,.508,.703}
\definecolor{navy}{rgb}{0,0,.5}
\definecolor{LightCyan}{rgb}{.875,1,1}
\definecolor{DarkRed}{rgb}{.543,0,0}
\definecolor{HotPink}{rgb}{1,.41,.70}
\definecolor{ForestGreen}{rgb}{.13,.54,.13}
\definecolor{OliveDrab}{rgb}{.42,.55,.14}
\definecolor{MediumBlue}{rgb}{0,0,.80}
\definecolor{RoyalBlue}{rgb}{.25,.41,.88}
\definecolor{DeepSkyBlue}{rgb}{0,.746,1}
\definecolor{Brown}{rgb}{0.545,0.271,0.074}
\definecolor{Purple}{rgb}{0.637,0.285,0.641}

\def\bea{\begin{eqnarray}}
\def\eea{\end{eqnarray}}
\def\bec{\begin{center}}
\def\ec{\end{center}}

\def\beq{\begin{equation}}
\def\eeq{\end{equation}}

\def\f{\frac}
\newcommand\lsim{\mathrel{\rlap{\lower4pt\hbox{\hskip1pt$\sim$}}
    \raise1pt\hbox{$<$}}}
\newcommand\gsim{\mathrel{\rlap{\lower4pt\hbox{\hskip1pt$\sim$}}
    \raise1pt\hbox{$>$}}}
\def\bea{\begin{eqnarray}}
\def\eea{\end{eqnarray}}
\def\ba{\begin{array}}
\def\ea{\end{array}}
\def\bc{\begin{center}}
\def\ec{\end{center}}
\def\nn{\nonumber}

\def\f{\frac}

\def\f#1#2{\frac{#1}{#2}}

\newcommand{\red}[1]{\textcolor{black}{#1}} 
\newcommand{\blue}[1]{\textcolor{black}{#1}} 

\begin{document}

%

\title{\Large Model Independent Constraints on Charges of New Particles}

\author{Dongjin Chway}
\email{djchway@gmail.com}
\affiliation{Department of Physics and Astronomy
and Center for Theoretical Physics, Seoul National University, Seoul 08826, Korea}
\affiliation{Center for Theoretical Physics of the Universe, Institute for Basic Science (IBS), Daejeon, 34051, Korea}
\author{Radovan Derm\'i\v{s}ek}
\email{dermisek@indiana.edu}
\affiliation{Department of Physics and Astronomy
and Center for Theoretical Physics, Seoul National University, Seoul 08826, Korea}
\affiliation{Physics Department, Indiana University, Bloomington, IN 47405, USA}
\author{Tae Hyun Jung}
\email{thjung0720@gmail.com}
\affiliation{Center for Theoretical Physics of the Universe, Institute for Basic Science (IBS), Daejeon, 34051, Korea}
\author{Hyung Do Kim}
\email{hdkim@phya.snu.ac.kr}
\affiliation{Department of Physics and Astronomy
and Center for Theoretical Physics, Seoul National University, Seoul 08826, Korea}

\begin{abstract}

Any particle that is charged under $SU(3)_C$ and $U(1)_{EM}$ can mediate the $gg \rightarrow \gamma\gamma$ process through loops.
Near the threshold for the new particle pair production, gauge boson exchanges necessitate the resummation of ladder diagrams. 
We discuss the leading log order matching of the one-loop result with non-relativistic effective theory resummed result.
We show how the diphoton invariant mass spectrum varies depending on decay width, color representation and electric charge of the new particle. The exclusion limits on the product of $SU(3)_C$ and $U(1)_{EM}$ charges
of the new scalar or fermion particle are obtained from current LHC data. 

\end{abstract}

\maketitle

\tableofcontents

\noindent
\section{Introduction} 

Direct production of new particles typically provides the best opportunities to search for them. However, in principle, if the decay modes of the new particle are complicated, involving soft particles, missing energy, or a number of final states, it may be difficult to see the particle directly. Still, even in those cases, the new particle leaves imprints in collider experiments.

In the previous letter \cite{Chway:2015lzg}, we showed that any particle carrying $SU(3)_C$ and $U(1)_{\rm EM}$ charges can mediate the $gg\rightarrow\gamma\gamma$ process through loops.
We obtained the constraints on the combined charge (which is the product of $SU(3)_C$ and $U(1)_{\rm EM}$ charges) in large charge limit when interference with the standard model quarks can be neglected. Near the threshold for the new particle pair production, gauge boson exchanges necessitate the resummation of ladder diagrams. In this paper, we present detailed explanation of the threshold resummation and the leading log order matching of the one-loop result with non-relativistic effective theory resummed result. 
We show how the diphoton invariant mass spectrum varies depending on decay width, color representation and electric charge of the new particle. 
We also include interference with the standard model quarks which is important for new particles with small combined charges. 
Finally we present new exclusion limits from current LHC data.

The larger the charges, the bigger their effects on the $gg\rightarrow\gamma\gamma$ cross section. At energies far above the threshold of the new particle pair production, it will give extra contribution to the cross section. However, around the threshold, it will provide a characteristic signal shape due to threshold physics. If the particle has a small decay width, the particle anti-particle pair will form bound states and show clear bound state resonances in the diphoton invariant mass spectrum. However, even if the particle has a large decay width and does not form clear bound states, it changes the shape of the diphoton spectrum.

As the decay width increases, the resonances of bound states are smeared and one cannot apply spectroscopic approach developed for charmonium and bottomonium. Instead, we can use methods developed for toponium. In their pioneering papers \cite{Fadin:1987wz, Fadin:1988fn}, Fadin and Khoze proposed how to treat top quark anti-top quark pair production in threshold region when top quark has a large decay width. Strassler and Peskin \cite{Strassler:1990nw} provided 
more clear explanation.
Threshold resummation effects in $t\bar{t}$ production with $b\bar{b}W^+W^-$ in the final state were further studied in Ref. \cite{Sumino:1992ai}.
Diphoton final state was studied for a new particle with a small decay width where the narrow width approximation is valid \cite{Kats:2009bv, Kats:2012ym}. 

A smaller decay width of a new particle, X, results in a larger branching ratio of X-onium into diphotons and thus leads to bigger resonances. In order to obtain a conservative bound, we needed a formalism applicable to large widths. Furthermore, smaller signal means that interference with the standard model process becomes important. Therefore, cross sections from the narrow width approximation are not sufficient and we needed to obtain amplitudes.
In the Higgs study on threshold effects in diphoton final states \cite{Melnikov:1994jb}, Melnikov, Spira, and Yakovlev already dealt with the same problem but only for triangular Feynman diagrams. 
Within the concept of non-relativistic effective field theory \cite{Caswell:1985ui}, we review their method and apply it to the diphoton process where not only triangular diagrams but also box and bubble diagrams appear as in Figs. \ref{diagrama} and \ref{diagramas}.
Additionally, they treated renormalization scale ambiguity appearing in the leading log of non-relativistic terms by comparing them with two-loop result.
Relying on the effective field theory, we suggest a prescription to keep the leading log order without knowing two-loop result.

The exclusion limits on the combined $SU(3)_C$ and $U(1)_{\rm EM}$ charge we present are independent of and often stronger than existing limits on separate charges obtained from other processes.
For example, the bounds were obtained from the fact that a new charged particle changes the running of the corresponding coupling. From Drell-Yan process, constraints were obtained on particles with electroweak charges \cite{Gross:2016ioi, Goertz:2016iwa}. From the ratio of 3 to 2 jets cross section, constraints were obtained on color charged new particles \cite{Becciolini:2014lya}.

This paper is organized as follows. In section \ref{ThresholdResummation}, we briefly review how getting an amplitude corresponds to solving a Schroedinger equation. We explain how to treat the renormalization scale appearing in Green's function which is the solution to the Schroedinger equation in section \ref{Methodology}. In section \ref{amplitudeshapesection}, amplitude shapes in the leading order and the leading log approximation of the Green's functions are compared. In section \ref{fermionshape} and \ref{signalshapesscalar}, we show signal shapes with a variety of new particle properties: decay width, color factor, combined charge, and electric charge. In section \ref{exclusionsection}, the exclusion limits on the combined charge of a new particle are updated, including the interference effects, using current LHC data.

\section{Threshold Resummation}
\label{ThresholdResummation}

\subsection{Threshold Singularities}
\label{ThresholdSingularities}

\begin{figure}[t]
\begin{center}
\begin{subfigure}[b]{0.23\textwidth}
\includegraphics[width=\textwidth]{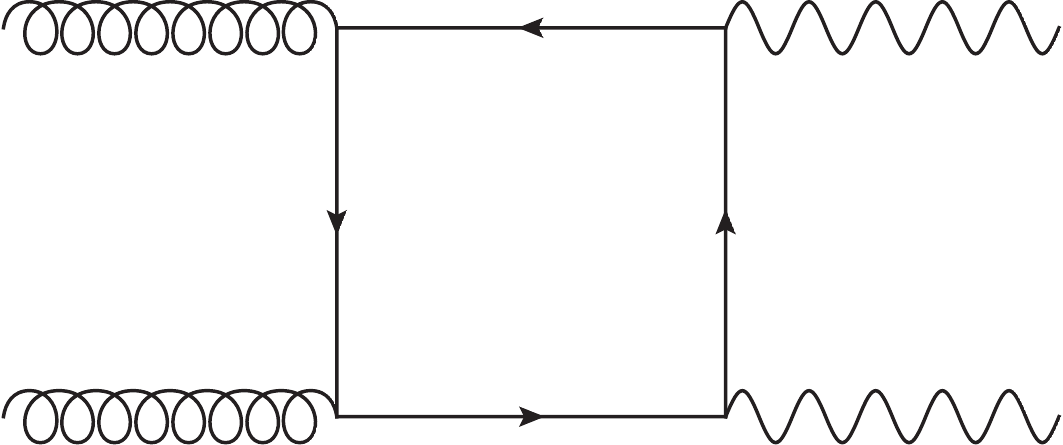}
\caption{}
\label{diagrama}
\end{subfigure}
\begin{subfigure}[b]{0.23\textwidth}
\includegraphics[width=\textwidth]{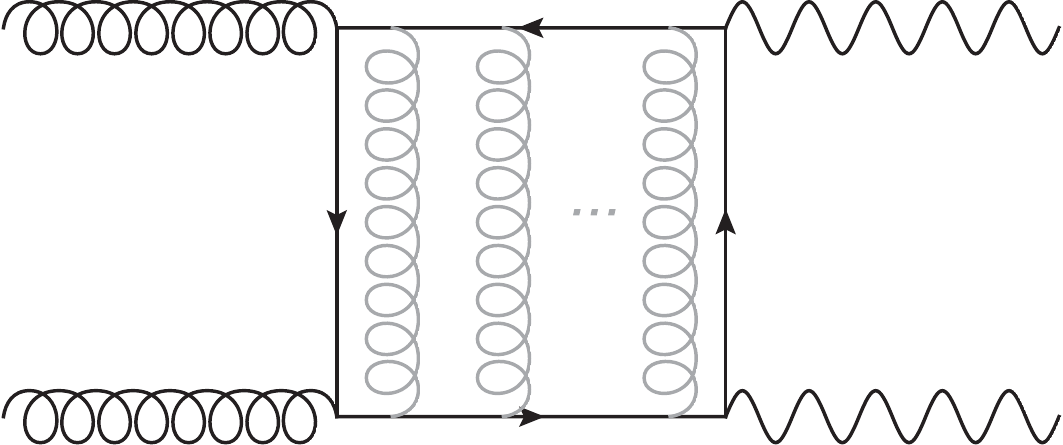}
\caption{}
\label{diagramb}
\end{subfigure}
\caption{Feynman diagrams for $gg \rightarrow \gamma \gamma$ mediated by a fermion with no gluon exchange (a) and with ladder gluon exchanges (b). Twisted topologies are not shown.
}
\label{diagram}
\end{center}

\begin{center}
\begin{subfigure}[b]{0.46\textwidth}
\includegraphics[width=\textwidth]{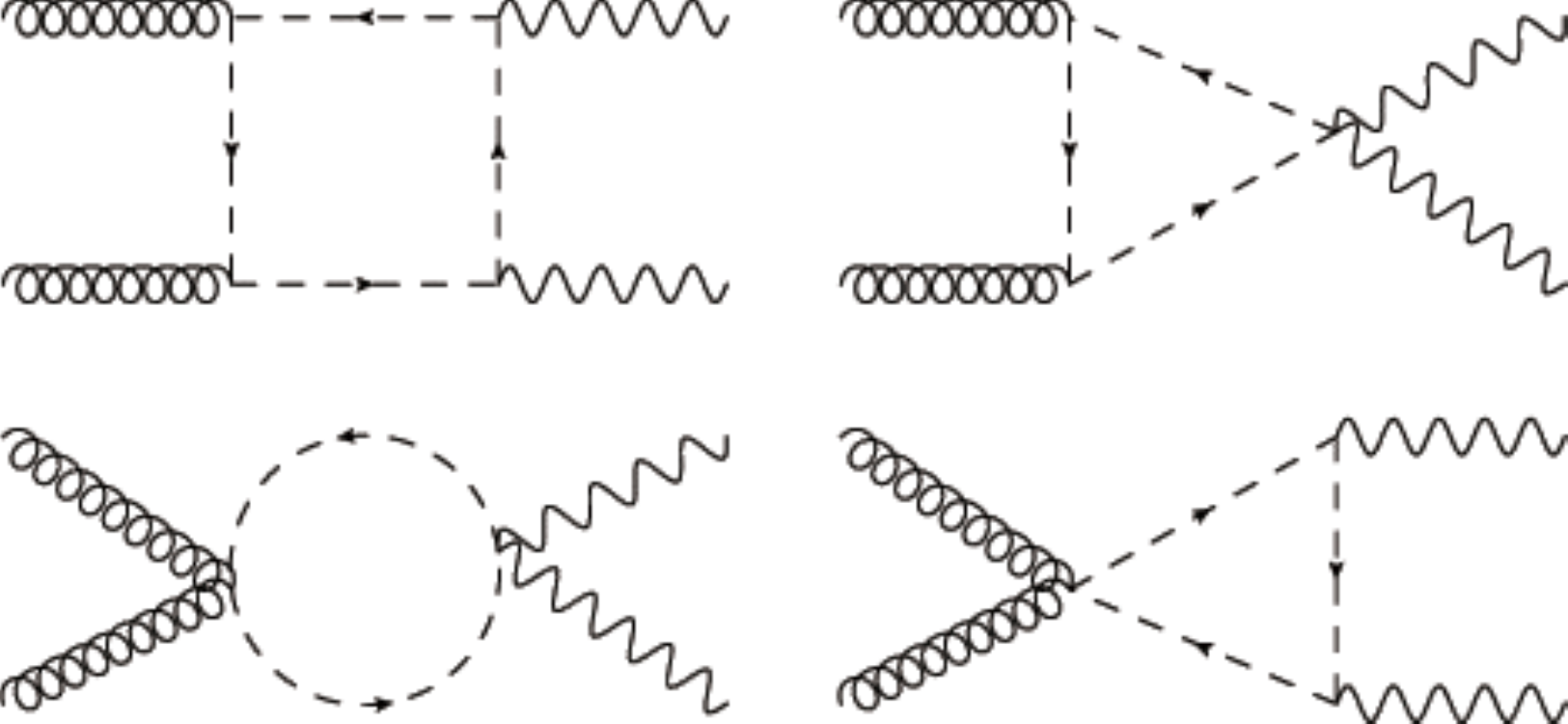}
\caption{}
\label{diagramas}
\end{subfigure}
\begin{subfigure}[b]{0.46\textwidth}
\includegraphics[width=\textwidth]{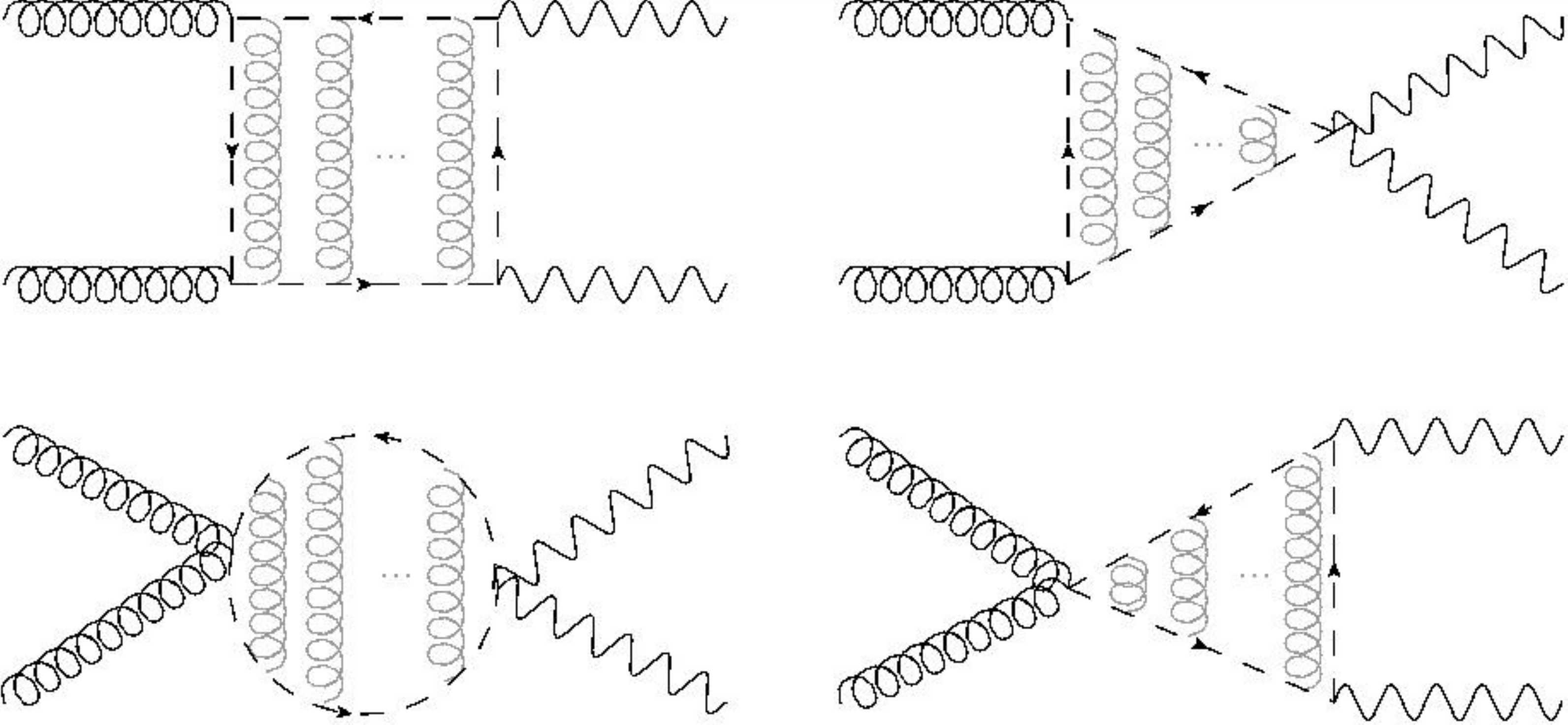}
\caption{}
\label{diagrambs}
\end{subfigure}
\caption{Feynman diagrams for $gg \rightarrow \gamma \gamma$ mediated by a scalar with no gluon exchange (a) and with ladder gluon exchanges (b). Twisted topologies are not shown.
}
\label{diagrams}
\end{center}
\end{figure}

\begin{figure}[t]
\includegraphics[width=0.8\textwidth]{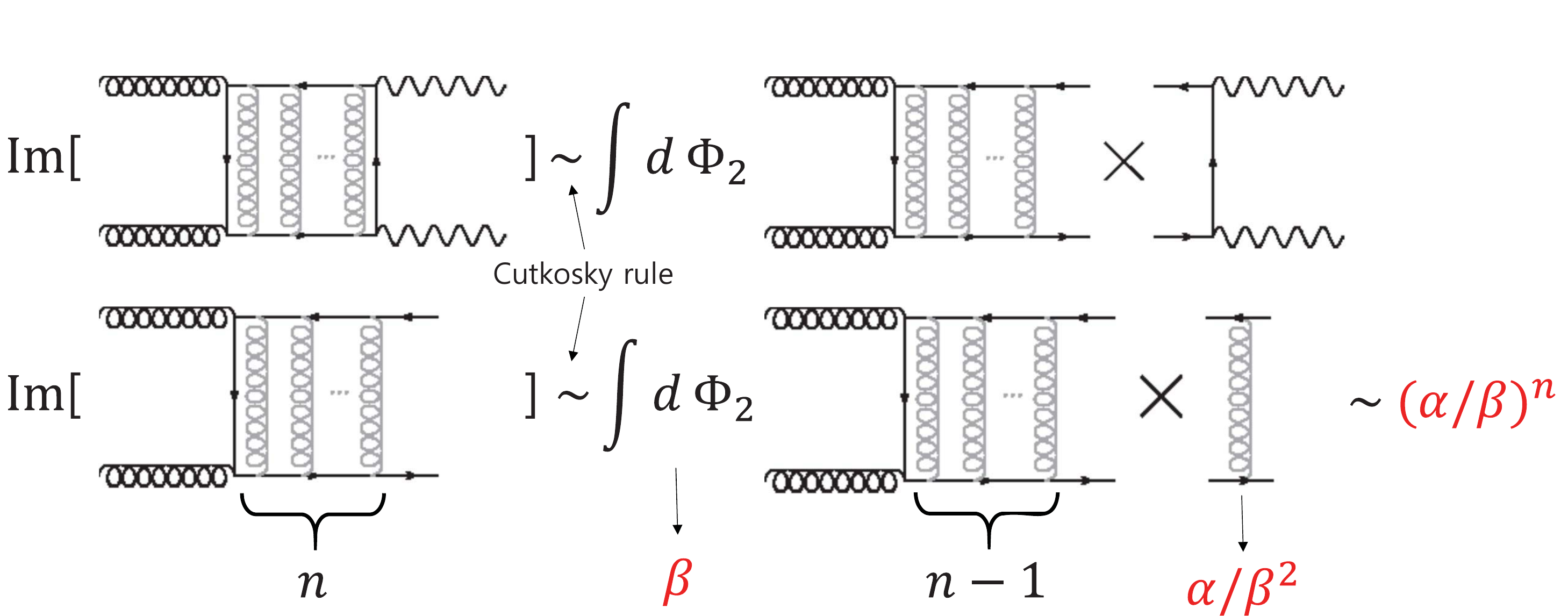}
\caption{Cutting rules to illustrate the appearance of factors of $\alpha$ and $\beta$ for the amplitude with $n$ exchanged gluons.}
\label{cutting}
\end{figure}

Model independently, any particle that is charged under $SU(3)_C$ and $U(1)_{EM}$ can mediate $gg \rightarrow \gamma \gamma$ process through loops, see Fig. \ref{diagrama} for a fermion or Fig. \ref{diagramas} for a scalar. These loops are proportional to the
combined charge of a particle X \cite{Chway:2015lzg}
\bea
C_X = N_X T_{R_X} Q_X^2,
\eea
where $T_{R_X}$ and $Q_X$ are its Dynkin index of $SU(3)_C$ representation and electric charge. $N_F$ is the number of copies of Dirac fermions in case X is a fermion and $N_S$ is the number of copies of complex scalars in case X is a scalar.
However, near the threshold of the loop-particle pair production, expansion in the usual loop number counting breaks down and the one-loop result is not a good approximation.
This can be schematically seen using the cutting rules \cite{Cutkosky:1960sp} as illustrated in Fig. \ref{cutting}.

\begin{figure}[t]
\includegraphics[width=0.65\textwidth]{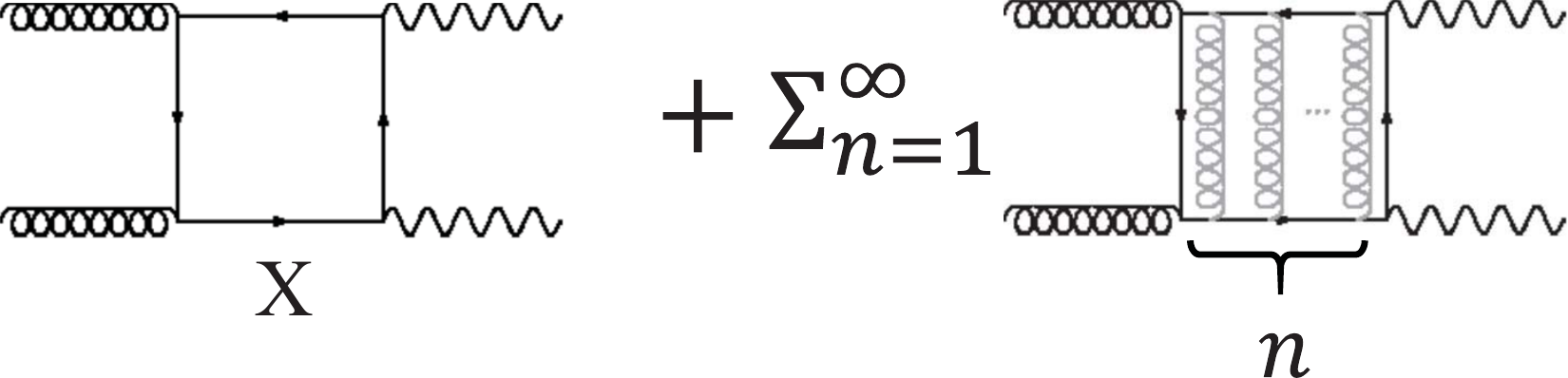}
\caption{Box diagram resummation over the number of ladder gluon exchanges.}
\label{BoxResum}
\end{figure}

In Fig. \ref{cutting}, the full diagram is cut into simpler parts.
For each cut, we obtain a factor of the velocity of the loop particle X, $\beta$, from the two particle phase space volume, $\int d \Phi_2$.
For each t-channel gauge particle exchange between nearly on-shell loop particles, we obtain $\alpha_s / \beta^2$.
\red{Applying the cutting rule to the zero gluon exchanged diagrams in Fig. \ref{diagrama} and Fig. \ref{diagramas}, we see that their imaginary parts start with $\beta^1$.
On the other hand, their real parts can start with $\beta^0$.}
\footnote{\red{The real part of the one loop box is actually small at threshold, but this has nothing to do with the power of $\beta$. It is accidentally small.}}
\red{Applying the cutting rule recursively, diagrams with $n$ exchanged gluons in Fig. \ref{diagramb} and Fig. \ref{diagrambs} are thought to have terms with an extra factor of $\alpha_s^n/\beta^{n-1}$ compared to the real part of no gluon exchange amplitude.} 
\footnote{\red{The cutting rule relates the $n-1$ gluon exchange amplitude and only the imaginary part of the $n$ gluon exchange amplitude. Therefore, in order to see the appearance of $\alpha_s^n/\beta^{n-1}$ rigorously including the real parts, one has to look at a recursive relation like Eq. \eqref{selfconeq} which relates complex amplitudes.}}
As the total energy gets closer to the threshold energy, $2 m_X$, $\beta$ goes to zero and infinities appear. Therefore, we have to sum all ladder diagrams as shown in Fig. \ref{BoxResum} for a fermion and similarly for a scalar.

There are other diagrams which have the same order in $\alpha_s$ expansion, but after the resummation, those diagrams will be higher order in $\alpha_s$ or $\beta$. 
For example, $n$-gluon exchange diagram with two exchanges crossed gives $\alpha_s^n/\beta^{n-2}$. Another example is that if one of the ladder gluon exchanges has three gluon self interaction, then one can think of it as a higher order correction to $\alpha_s$ after the resummation. 

\subsection{Non-Relativistic Effective Field Theory}

\begin{figure}[t]
\includegraphics[width=0.8\textwidth]{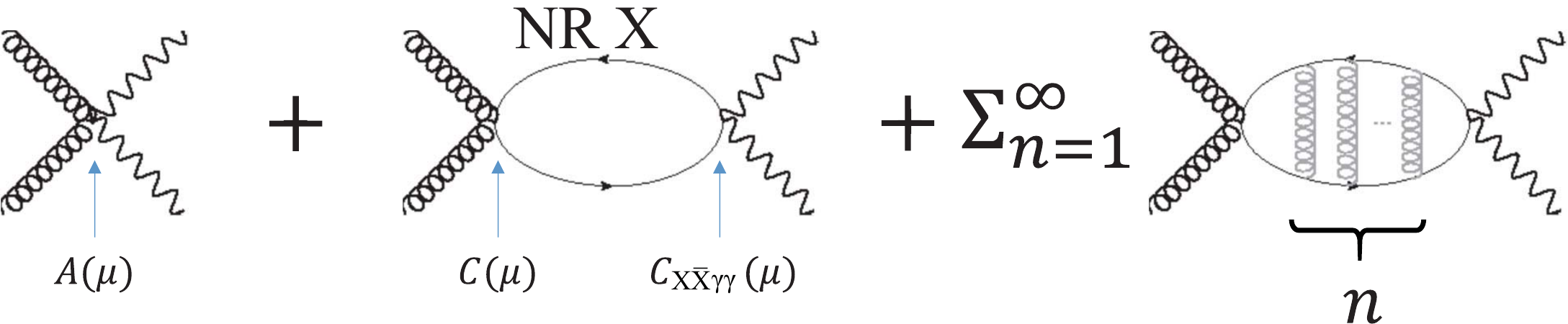}
\caption{The resummation of ladder gluon exchanges for $gg\rightarrow\gamma\gamma$ mediated by a fermion in the non-relativistic effective theory.}
\label{EffResum}
\end{figure}

The resummation of the ladder diagrams in Fig. \ref{BoxResum} can be performed in the effective field theory where the relativistic part of the loop particle X is integrated out. In that theory, we now have $gg\gamma\gamma$ vertex which is absent in the full theory. Let us for the moment suppress polarization indices and call it $A(\mu)$, where $\mu$ is a renormalization scale.
We also obtain $ggX\bar{X}$ vertex which we call $C$ and $\gamma\gamma X\bar{X}$ vertex which we call $C_{X \bar{X} \gamma \gamma}$. Now $X$ and $\bar{X}$ are non-relativistic particles and thus non-relativistic propagators should be used for them. 
The resummation of the box diagram in full theory, Fig. \ref{BoxResum}, corresponds to the resummation in the non-relativistic effective theory shown in Fig. \ref{EffResum}.

Non-relativistic fermion and scalar propagators are:
\bea
S\left(m_X+\f{E}{2}+k^0,\vec{k}\right)\simeq \f{\f{i\left(1 \pm \gamma^0\right)}{2}}{k^0+\f{E}{2}-\f{\vec{k}^2}{2m_X}}, \label{eq2}\\
D\left(m_X+\f{E}{2}+k^0,\vec{k}\right)\simeq \f{i}{k^0+\f{E}{2}-\f{\vec{k}^2}{2m_X}}.
\eea 
In the numerator of the fermion propagator, the plus or minus sign is for fermion or anti-fermion, respectively. For simplicity, in this section we present formulas only for a scalar. In order to obtain the corresponding formulas for a fermion, one needs to keep numerators in Eq. \eqref{eq2}, see also Ref. \cite{Strassler:1990nw}.  

\begin{figure}[t]
\includegraphics[width=0.30\textwidth]{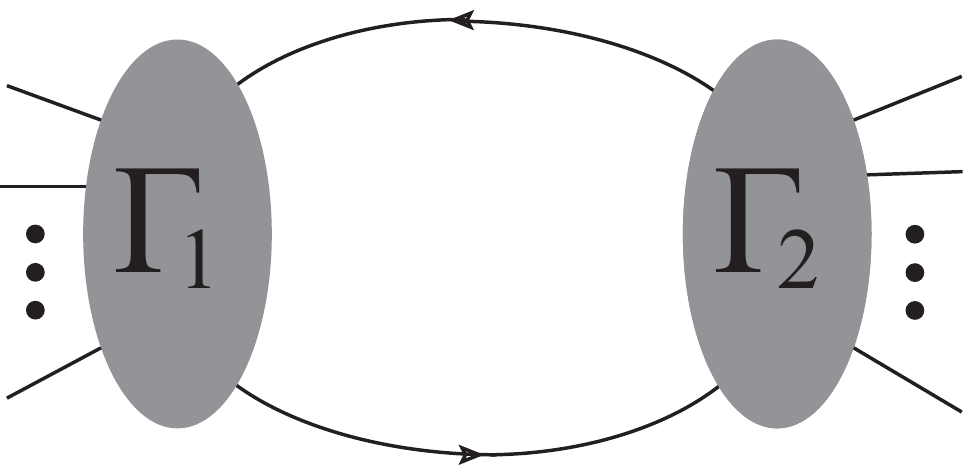}
\caption{A loop with a pair of non-relativistic $X$ and $\bar{X}$ attached to $\Gamma_1$ and $\Gamma_2$.}
\label{nrloop}
\end{figure}

Counting divergences in this effective theory is a bit tricky.
A loop containing a pair of $X$ and $\bar{X}$, which is shown in Fig. \ref{nrloop}, gives
\bea
&&\int \f{d^4 k}{(2 \pi)^4} \Gamma_1 \f{i}{k^0+\f{E}{2}-\f{\vec{k}^2}{2m_X}} \Gamma_2 \f{i}{-k^0+\f{E}{2}-\f{\vec{k}^2}{2m_X}}  \\
&=&  \int \f{d^3 k}{(2 \pi)^3} \Gamma_1 \f{i}{E-\f{\vec{k}^2}{m_X}} \Gamma_2 \\
&=&  \int \f{d^3 k}{(2 \pi)^3} \Gamma_1 \left(-i\right) \tilde{G}_0(E,\vec{k}) \Gamma_2 
\eea
after evaluating a contour integral over $k^0$.
Note that $\tilde{G}_0$ is the momentum space Green's function solution to the Schroedinger equation with no potential. 
We see that, in divergence counting, a loop integration gives +3 powers of momenta and a pair of non-relativistic propagators gives -2. 
Plus, a gauge particle exchange between two non-relativistic particles gives -2.

\begin{figure}[t]
\includegraphics[width=0.6\textwidth]{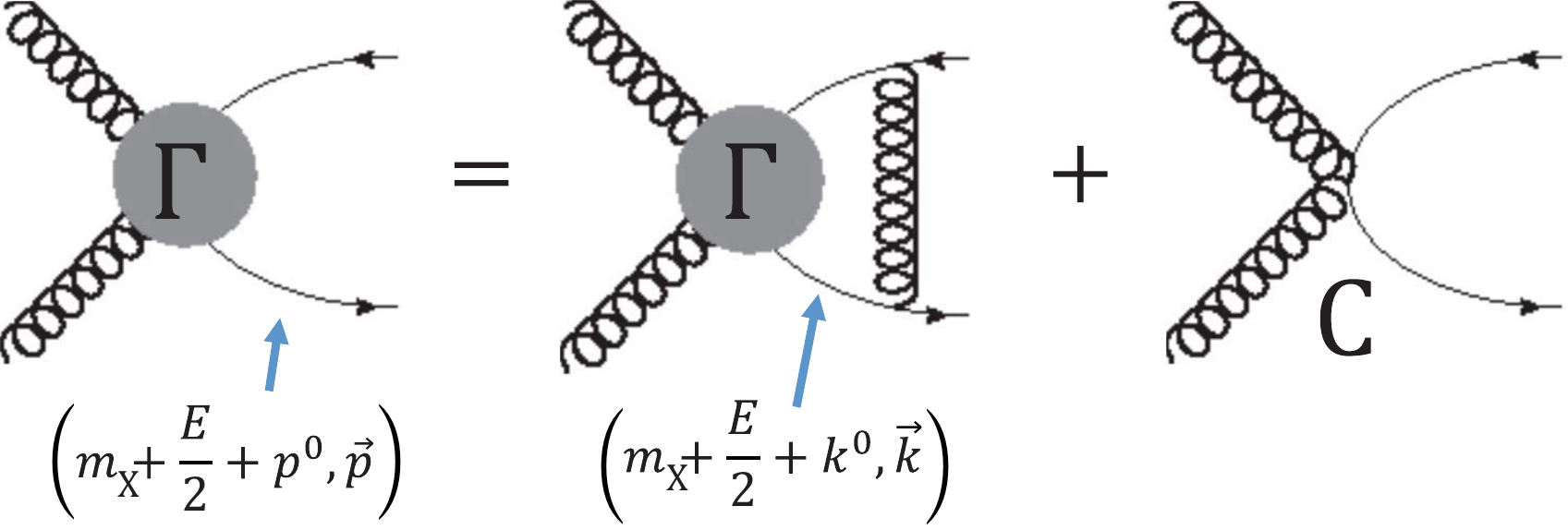}
\caption{Self consistency condition of the gluon exchange resummed vertex, $\Gamma$. The resummed vertex should be the same as the sum of the at least one gluon exchanged part and the no gluon exchanged vertex, $C$. The outgoing particle momentum is $(m_x+\f{E}{2}+p^0,\vec{p})$ and the loop particle momentum is $(m_x+\f{E}{2}+k^0,\vec{k})$. The loop integration is over $(k^0,\vec{k})$.}
\label{selfcon}
\end{figure}
Let us call the gluon exchange resummed $ggX\bar{X}$ vertex $\Gamma$. 
Then, Strassler and Peskin showed in Ref. \cite{Strassler:1990nw} that it should satisfy a self-consistency condition shown in Fig. \ref{selfcon}. The $\Gamma$ can be separated into two parts: at least one gluon exchange part and no gluon exchange part,
\bea
\Gamma(E,\vec{p})=C+\int \f{d^3 k}{(2\pi)^3} \Gamma(E,\vec{k}) \tilde{G}_0(E,\vec{k}) \f{C_C g_s^2}{(\vec{p}-\vec{k})^2},
\label{selfconeq}
\eea
where $C$ is the tree level of $g g X \bar{X}$ vertex and the color factor in Coulomb potential, $C_C$, is given by $C_C=C_{2,X}-\f{1}{2}C_{2,X\bar{X}}$, where $C_{2,X}$ is the quadratic Casimir for the particle X and $C_{2,X\bar{X}}$ is that for the bound state \cite{Kats:2009bv, Kats:2012ym}. For example, the $C_{2,X}$ is $4/3$ for the particle X in the fundamental representation, the $C_{2,X\bar{X}}$ is $0$ for the singlet bound state, and thus $C_C$ is $4/3$. The next smallest $C_{2,X}$ is $3$ and it is for the particle X in the octet representation.
For the $gg\rightarrow\gamma\gamma$ process, the $C_{2,X\bar{X}}$ should be zero. For the $gg\rightarrow g\gamma$ process, the $C_{2,X\bar{X}}$ should be $3$. In this process, the $C_C$ is $-1/6$ for the particle X in the fundamental representation. The $gg\rightarrow gg$ process opens more possibilities. The $C_C=-1$ is the negative number of biggest magnitude, which is obtained for the particle in the octet representation and the bound state in the representation of dimension $\textbf{27}$.

One can manipulate the equation by multiplying it with $\tilde{G}_0(E,\vec{p})/C$ on both sides.
\bea
\f{\tilde{G}_0(E,\vec{p})}{C}\Gamma(E,\vec{p}) = \tilde{G}_0(E,\vec{p})+\tilde{G}_0(E,\vec{p}) \int \f{d^3 k}{(2\pi)^3} \Gamma(E,\vec{k}) \f{\tilde{G}_0(E,\vec{k})}{C} \f{C_C g_s^2}{(\vec{p}-\vec{k})^2}.
\eea
Defining
\bea
\tilde{G}(E,\vec{p}) \equiv \f{\tilde{G}_0(E,\vec{p})}{C}\Gamma(E,\vec{p}),
\label{Gdef}
\eea
we obtain a very familiar form,
\bea
\tilde{G}(E,\vec{p}) = \tilde{G}_0(E,\vec{p})+\tilde{G}_0(E,\vec{p}) \int \f{d^3 k}{(2\pi)^3} \tilde{G}(E,\vec{k}) \f{C_C g_s^2}{(\vec{p}-\vec{k})^2}.
\eea
Therefore, $\tilde{G}(E,\vec{k})$ is the Fourier transform of $G(E,\vec{x})$ that satisfies the Schroedinger equation with Coulomb potential,
\bea
\left(-\f{\nabla^2}{m_X}-\f{C_C \alpha_s}{r}-E\right) G(E,\vec{x}) = \delta^3(\vec{x}).
\label{schroedinger}
\eea

\begin{figure}[t]
\includegraphics[width=0.8\textwidth]{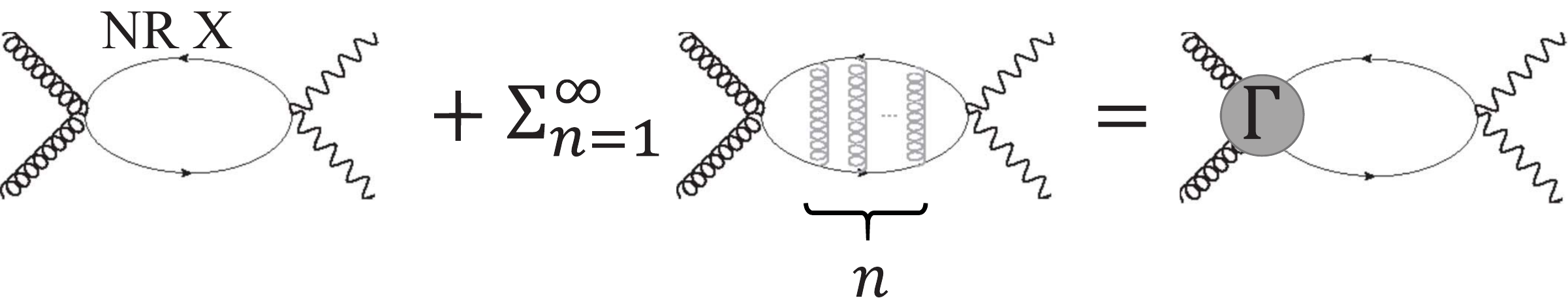}
\caption{The gluon exchange resummation represented by the gluon exchange resummed vertex, $\Gamma$.}
\label{resumother}
\end{figure}
Melnikov, Spira, and Yakovlev noticed in Ref. \cite{Melnikov:1994jb} that the resummation over the number of gluon exchanges in loop diagrams can be represented by using the vertex $\Gamma$. Fig. \ref{resumother} shows how they are related.
The amplitude of the diagram using the resummed vertex is
\bea
&&-\int \f{d^3 p}{(2\pi)^3} \Gamma(E,\vec{p}) \tilde{G}_0(E,\vec{p}) C_{X\bar{X}\gamma\gamma} \\
&=& - \int \f{d^3 p}{(2\pi)^3} \f{C \tilde{G}(E,\vec{p})}{\tilde{G}_0(E,\vec{p})} \tilde{G}_0(E,\vec{p}) C_{X\bar{X}\gamma\gamma} \\
&=& - C C_{X\bar{X}\gamma\gamma} G(E,\vec{x}=\vec{0}),
\eea
where Eq. \eqref{Gdef} is used in the first identity.
Thus, including the tree level vertex, the complete resummed result is
\bea
A(\mu) - C C_{X\bar{X}\gamma\gamma} G(E,\vec{x}=\vec{0}).
\label{resummedresult}
\eea

The Green's function is well known in an analytic form,
{\small\bea
&&  G(E,\vec{x}=\vec{0})    \label{green} \\
&&  = - \frac{m_X^2}{4\pi} \left\{ \sqrt{-\frac{E}{m_X}-i \epsilon} 
 -C_C \alpha_s(\mu) \ln \left(\mu\sqrt{\frac{1}{-m_X E}+i \epsilon}\right)
  -\frac{2}{\sqrt{m_X}}\sum_{n=1}^{\infty}\frac{E_n}{\sqrt{(-E-i \epsilon)}-\text{sign}(C_C)\sqrt{E_n} }\right\}  \nn,
\eea}
where $E_n=C_C^2 \alpha_s^2 m_X/4n^2$.
Note that the Green's function has a $\mu$ dependence coming from $\ln (\f{m_X\beta}{\mu})$.
Here, $\beta \simeq \sqrt{|E|/m_X}$ is the velocity of the loop particle X.

\subsection{Renormalization Scale}

\begin{figure}[t]
\begin{subfigure}[b]{0.115\textwidth}
\includegraphics[width=\textwidth]{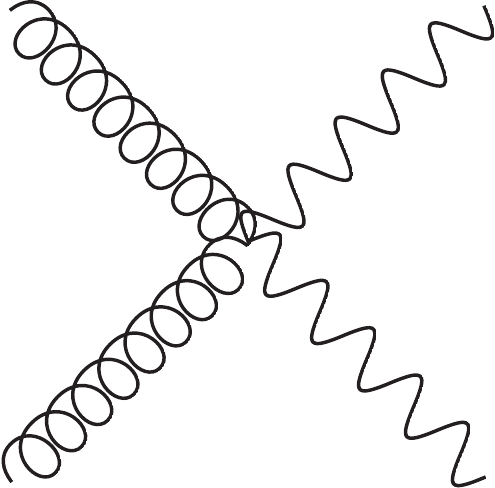}
\caption{}
\label{Effver1}
\end{subfigure}
\begin{subfigure}[b]{0.25\textwidth}
\includegraphics[width=\textwidth]{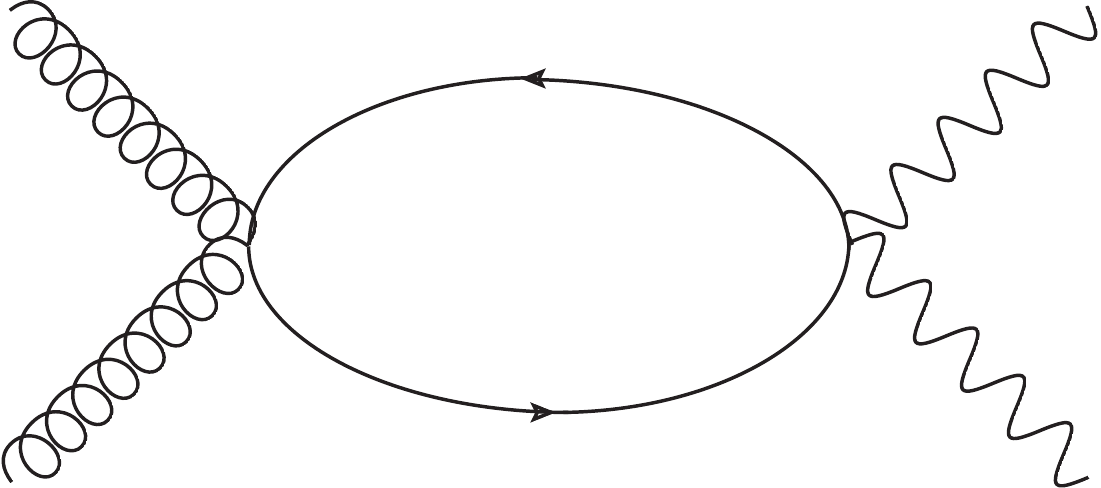}
\caption{}
\label{Effver2}
\end{subfigure}
\begin{subfigure}[b]{0.25\textwidth}
\includegraphics[width=\textwidth]{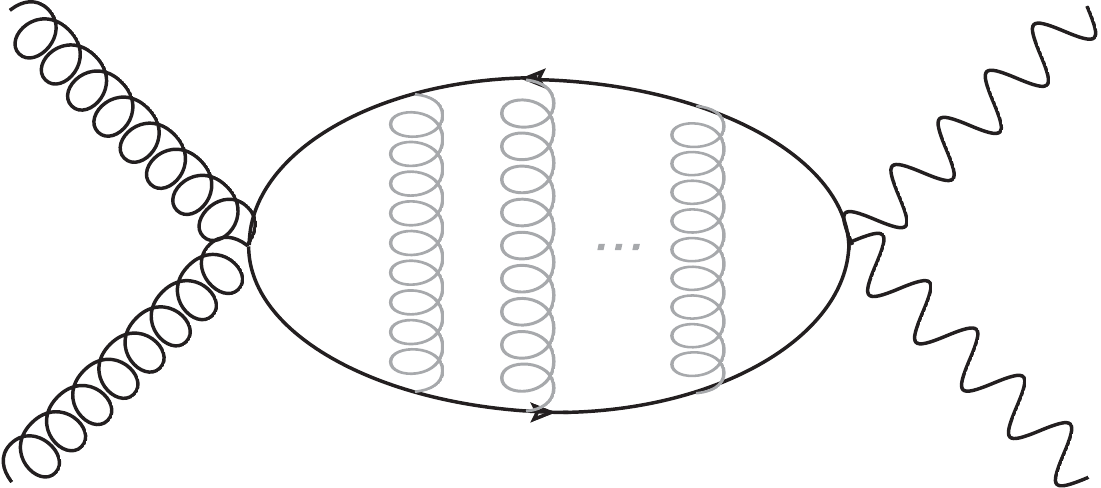}
\caption{}
\label{Effver3}
\end{subfigure}
\caption{From an effective field theory where relativistic part of top and anti-top is integrated out, Feynman diagrams for $gg \rightarrow \gamma \gamma$ with tree level (a), no gluon exchange (b) and gluon exchanges (c). Grey gluons indicate ladder diagrams.}
\label{diagrameff}
\end{figure}
To see the origin of the $\mu$ dependence, let us go back to amplitudes before the resummation.
The amplitude of the n-gluon exchanged diagram given in Fig. \ref{Effver3} is proportional to
\bea
\int \f{d^3 k_1}{(2\pi)^3} \f{d^3 k_2}{(2\pi)^3} ... \f{d^3 k_{n+1}}{(2\pi)^3}
\f{1}{E-\f{\vec{k}_1^2}{m_X}}\f{C_C g_s^2}{(\vec{k}_1-\vec{k}_2)^2}
\f{1}{E-\f{\vec{k}_2^2}{m_X}} \nonumber \\ 
\times \f{C_C g_s^2}{(\vec{k}_2-\vec{k}_3)^2}\f{1}{E-\f{\vec{k}_3^2}{m_X}} \cdots
\f{C_C g_s^2}{(\vec{k}_{n}-\vec{k}_{n+1})^2}\f{1}{E-\f{\vec{k}_{n+1}^2}{m_X}}.
\eea
As explained earlier, we can count divergence and we find that the amplitude with $n$ gluon exchanges has a $1-n$ divergence. Thus, for $n \geq 2$, amplitudes are convergent. There is no pole and no $\ln \mu$ term in dimensional regularization.

For $n=0$, the above integration gives
\bea
\f{m_X}{4\pi}\sqrt{- m_X E}.
\eea
Note that from the counting, this is linearly divergent, but because the integrals are over three-momenta and we work in dimensional regularization, we do not obtain any pole.

For $n=1$, the above integration gives
\bea
\f{m_X^2}{4\pi} C_C \alpha_s \left\{\f{1}{2\epsilon}-1-\ln \left(\f{\sqrt{- m_X E}}{\mu}\right) \right\}.
\label{n1}
\eea
Here $\alpha_s = g_s^2 / 4\pi^2$. Note that the integration is the same as for the sunset diagram in 3 dimensions with zero external momentum \cite{Groote:1998wy}. 
Because of the $1/\epsilon$ pole, there should be a counter term to cancel it.
In the non-relativistic effective theory, we have $gg\gamma\gamma$ vertex, see Fig. \ref{Effver1}.

Because of $\ln (\f{m_X\beta}{\mu})$ in Eq. \eqref{n1}, we also have to take into account the running of the $gg\gamma\gamma$ vertex, $A(\mu)$. This running is what is called a soft running. (Hard running is $\ln (m_X / \mu)$ or $\ln (m_{\gamma \gamma} / \mu)$ for the diphoton invariant mass $m_{\gamma \gamma}$, and ultra soft running is $\ln (m_X\beta^2 / \mu)$.) On the other hand, $ggX\bar{X}$ and $\gamma\gamma X\bar{X}$ vertices do not have soft running in the first order of $\alpha_s$.
This means that Eq. \eqref{resummedresult} is $\mu$ independent if we consider only the soft running.
We choose $\mu = m_X \beta$. This choice corresponds to typical gluon momentum scale exchanged in the ladder diagrams, so it is expected to give renormalization group improved potential and Green's functions.

\subsection{Effects of Non-Zero Decay Width}

Fadin and Khoze proposed how to deal with unstable loop particle case in their pioneering papers \cite{Fadin:1987wz, Fadin:1988fn}. They suggested to replace $E$ by $E+i\Gamma_X$, where $\Gamma_X$ is the decay width of the loop particle X. Accordingly,
\bea
\mu=\sqrt{m_X\sqrt{E^2+\Gamma_X^2}}
\label{runningmu}
\eea
is now our choice. 
They noticed that this $\Gamma_X$ plays the role of IR cutoff.
It cuts off $\mu$ smaller than $m_X \sqrt{\f{\Gamma_X}{m_X}}$.
One can also see this from the uncertainty relation, $\f{1}{\Gamma_X} (m_X\beta^2) > 1$, and the fact that typical gluon momentum exchanged in ladder diagrams is $m_X \beta$. For the top quark, this IR cutoff is about 15 GeV.

Other than each constituent particle decaying, the bound state can decay through an annihilation of the $X\bar{X}$, with the decay rate of order $m_X \bar{\alpha}_s^3 \alpha_s^2$ for a digluon final state after the threshold resummation. Here, $\bar{\alpha}_s$ is the strong coupling evaluated at the soft scale, $m_X \bar{\alpha}_s$, and $\alpha_s$ is the strong coupling evaluated at hard scale, $m_X$. Tree level effective vertex before resummation is a four vertex from $X\bar{X} \rightarrow g g \rightarrow X\bar{X}$. One can follow the same procedure for this vertex to do the resummation. However, this is not of interest in this paper because the decay width of the bound state through annihilation is order of $10^{-5} m_X$ and we are going to discuss decay widths of the constituent particle larger than $10^{-4} m_X$.

\section{Methodology}
\label{Methodology}

\subsection{Matching}

The effective theory result, Eq. \eqref{resummedresult}, should be matched with the full theory result, Fig. \ref{BoxResum}, which we do not know how to sum.
Since the summation is only needed to take into account $\alpha_s/\beta$ expansion near the threshold, away from the threshold, one-loop (no gluon exchange) or two-loop (one gluon exchange) result gives good approximation.
As shown in the section \ref{ThresholdSingularities}, $n$ gluon exchanges result in $\alpha^n/\beta^{n-1}$, and thus for two-loop (one gluon exchange) we do not expect a threshold singularity of $1/\beta$. 
Therefore, we can assume that except for some possible large log terms, it is sufficient to keep the one-loop result, which was for fermion case already calculated for a light by light scattering \cite{Costantini:1971cj}. 
We obtained the fermion loop amplitudes and also scalar loop amplitudes in Veltman-Passarino basis integrals using FeynArts, FormCalc and LoopTools \cite{Hahn:2000kx}. 

In the effective theory, the amplitude of the single gluon exchange in Fig. \ref{Effver3} contains $C_C \alpha_s \ln (\sqrt{-m_X E})$, see Eq. \eqref{n1}.
At the same time, we should take into account the same kind of log term that comes from the one gluon exchange diagram in the full theory. General knowledge of effective field theory tells us that logs of low energy parameters should agree between a full theory and its effective theory. Thus, in order for the argument of the log to be dimensionless, the full theory must contain the term $\ln (\sqrt{-m_X E}/{m_X})$ with the same coefficient as in the effective theory. This can in principle be large for small $E$, and thus we should keep it in the matching. In other words, we are doing the leading log (LL) computation. By the way, from Eqs. \eqref{resummedresult} and \eqref{green}, we see that the log term shares its coefficient with $\sqrt{-E/m_X}$ term in the effective theory and so should they in the full theory. 

More explicitly, up to LL order, we take approximations
\bea
&&\mathcal{M}_{\rm UV 1-loop} (\sqrt{s}=2m_X+E) \simeq \mathcal{M}_{\rm UV 1-loop} (\sqrt{s}=2m_X) + B \sqrt{\frac{-E}{m_X}} ,  \\
&&\mathcal{M}_{\rm UV 2-loop} (\sqrt{s}=2m_X+E) \simeq  B C_C \alpha_s \ln \left(\f{\sqrt{-m_X E}}{m_X}\right),
\eea
where \red{the one-loop amplitude, $\mathcal{M}_{\rm UV 1-loop} (\sqrt{s}=2m_X+E)$, obtained in the full theory as a function of the diphoton invariant mass, $\sqrt{s}$, is expanded as a Taylor series about $\sqrt{E}=0$} and we ignore terms $\mathcal{O} \left( \alpha \alpha_s \left(\sqrt{\frac{-E}{m_X}}\right)^2\right)$, $\mathcal{O} \left(\alpha \alpha_s^2 \right)$, and $\mathcal{O} \left(\alpha^2 \alpha_s\right)$.
For each set of polarizations of initial gluons and final photons, this is matched to
\bea
A(\mu) \label{e12loop} + C C_{X\bar{X}\gamma\gamma} \frac{m_X^2}{4\pi} \left\{ \sqrt{-\frac{E}{m_X}} 
   -C_C \alpha_s(\mu) \ln \left(\frac{\mu}{\sqrt{-m_X E}}\right) \right\}.
\eea
Then, matching $\beta^1$ terms gives
\bea
 B =  C C_{X\bar{X}\gamma\gamma} \frac{m_X^2}{4\pi},
\eea
and the rest of the terms results in
\bea
\mathcal{M}_{\rm UV 1-loop} (\sqrt{s}=2m_X) = A(\mu) + B C_C \alpha_s(\mu) \ln \left(\f{m_X}{\mu}\right),
\eea
\red{where the logarithm can be interpreted as the renormalization group evolution of the effective $gg\gamma\gamma$ vertex $A(\mu)$ from the scale $\mu$ to the scale $m_X$.} 
Emphasizing again, this matching is to be done for each set of polarizations of initial gluons and final photons. Actually, $B$ is non-zero only for polarizations of $(++++)$, $(----)$, $(++--)$, and $(--++)$ , where the first two labels in parenthesis are the polarizations of initial state gluons and the last two are those of final state photons.
\red{Up to LL order, the matched effective vertices do not depend on the details of the model other than whether the new particle is a scalar or fermion. A model dependence will appear in higher order matching.}

Now, we can write the effective theory resummed result, Eq. \eqref{resummedresult}, using parameters obtained from the UV l-loop result, as
\bea
&& \mathcal{M}_{\rm UV 1-loop} (\sqrt{s}=2m_X) \nn \\
&&  + B \left\{ \sqrt{\frac{-E}{m_X}} -C_C \alpha_s(\mu) \ln \left(\f{m_X}{\sqrt{-m_X E}}\right) -\frac{2}{\sqrt{m_X}}\sum_{n=1}^{\infty}\frac{E_n}{\sqrt{-E}-\text{sign}(C_C)\sqrt{E_n}} \right\}
\eea
or, using the Green's function, as
\bea
\mathcal{M}_{\rm UV 1-loop} (\sqrt{s}=2m_X) &-& B C_C \alpha_s(\mu) \ln \left(\f{m_X }{\mu}\right) - B \f{4\pi}{m_X^2}G(E,\vec{0}) , \label{AmG}
\eea
where $G(E,\vec{0})$ is given in Eq. \eqref{green}.
In our calculations, keeping polarization indices and angular dependence,  we will use
\bea
\mathcal{M}_{\rm UV 1-loop}^{\lambda_1 \lambda_2 \lambda_3 \lambda_4} (s, t, u) - B^{\lambda_1 \lambda_2 \lambda_3 \lambda_4} \sqrt{\f{-(\sqrt{s}-2m_X)}{m_X}}  
- B^{\lambda_1 \lambda_2 \lambda_3 \lambda_4} C_C \alpha_s(\mu) \ln \left(\f{m_X }{\mu}\right)  \nn\\
- B^{\lambda_1 \lambda_2 \lambda_3 \lambda_4} \f{4\pi}{m_X^2}G(\sqrt{s}-2m_X,\vec{0})
 \label{WeUse} 
\eea
around the threshold.

\subsection{LL Green's Function}
The Green's function $G(E,\vec{0})$ sensitively depends on $\alpha_s(\mu)$ because it enters in the denominators in a pole like form, $\f{1}{\sqrt{-E} \pm \sqrt{E_n}}$. The $\alpha_s(\mu)$ is sensitive to the choice of $\mu$, and so $G(E,\vec{0})$ has a strong dependence on the unphysical parameter $\mu$. We want to cure it. Since the origin of $\alpha_s(\mu)$ dependence of $G(E,\vec{0})$ is from the gluon exchange (see Eqs. \eqref{selfconeq} and \eqref{resumother}), adding next order correction to it will reduce the $\mu$ dependence of $G(E,\vec{0})$. This is done by replacing the Coulomb potential, $-\f{C_C \alpha_s}{r}$, in Eq. \eqref{schroedinger} by
\bea
-\f{C_C \alpha_s(\mu)}{r}\left(1+\f{\alpha_s}{4\pi}\left( 2 \beta_0 \ln (\mu e^{\gamma_E} r ) + a_1 \right)\right),
\label{nlopot}
\eea
where $\beta_0 = 11- \f{2}{3} n_F$ and $a_1=\f{31}{3}-\f{10}{9}n_F$ for $n_F$ quarks lighter than gluon momentum exchange energy \cite{Fischler, Billoire}. The $\gamma_E$ is the Euler-Mascheroni constant. We can obtain $G(E,\vec{0})$ numerically as suggested in Ref. \cite{Kiyo:2010jm}. Let us call it $G_{\rm LL} (E,\vec{0})$. Large $\mu$ dependence disappears in the following expression,
\bea
\mathcal{M}_{\rm UV 1-loop} (\sqrt{s}=2m_X) &-& B C_C \alpha_s(\mu) \ln \left(\f{m_X }{\mu}\right) - B \f{4\pi}{m_X^2} G_{\rm LL}(E,\vec{0})  , \label{NLOmat}
\eea
and what we will use in our calculation is Eq. \eqref{WeUse} with $G_{\rm LL}$ in place of $G$.

\section{Amplitude Shapes: LO vs LL}
\label{amplitudeshapesection}

\begin{figure}[t]
\begin{subfigure}[b]{0.49\textwidth}
\includegraphics[width=\textwidth]{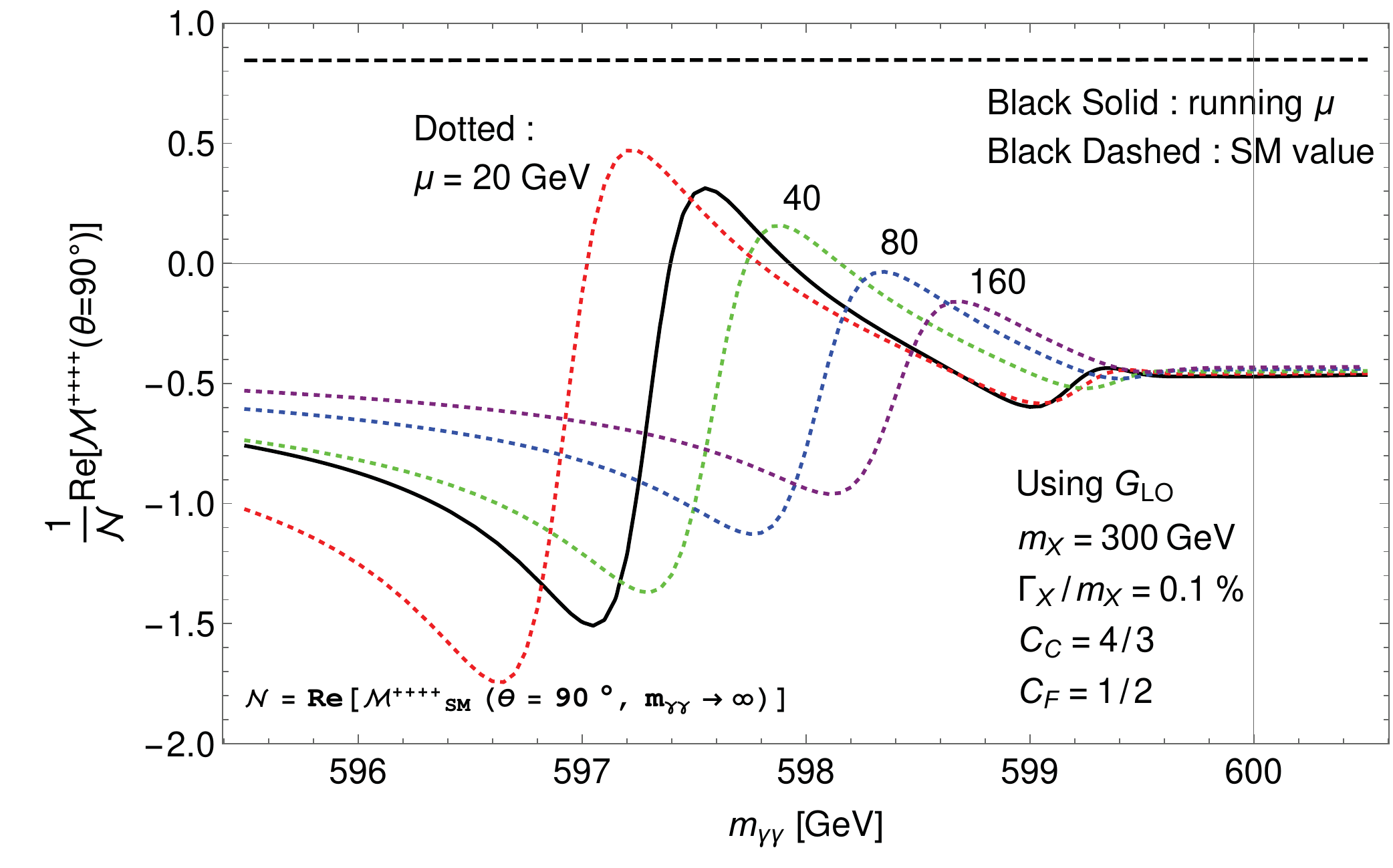}
\caption{}
\label{AmpLOreal}
\end{subfigure}
\begin{subfigure}[b]{0.49\textwidth}
\includegraphics[width=\textwidth]{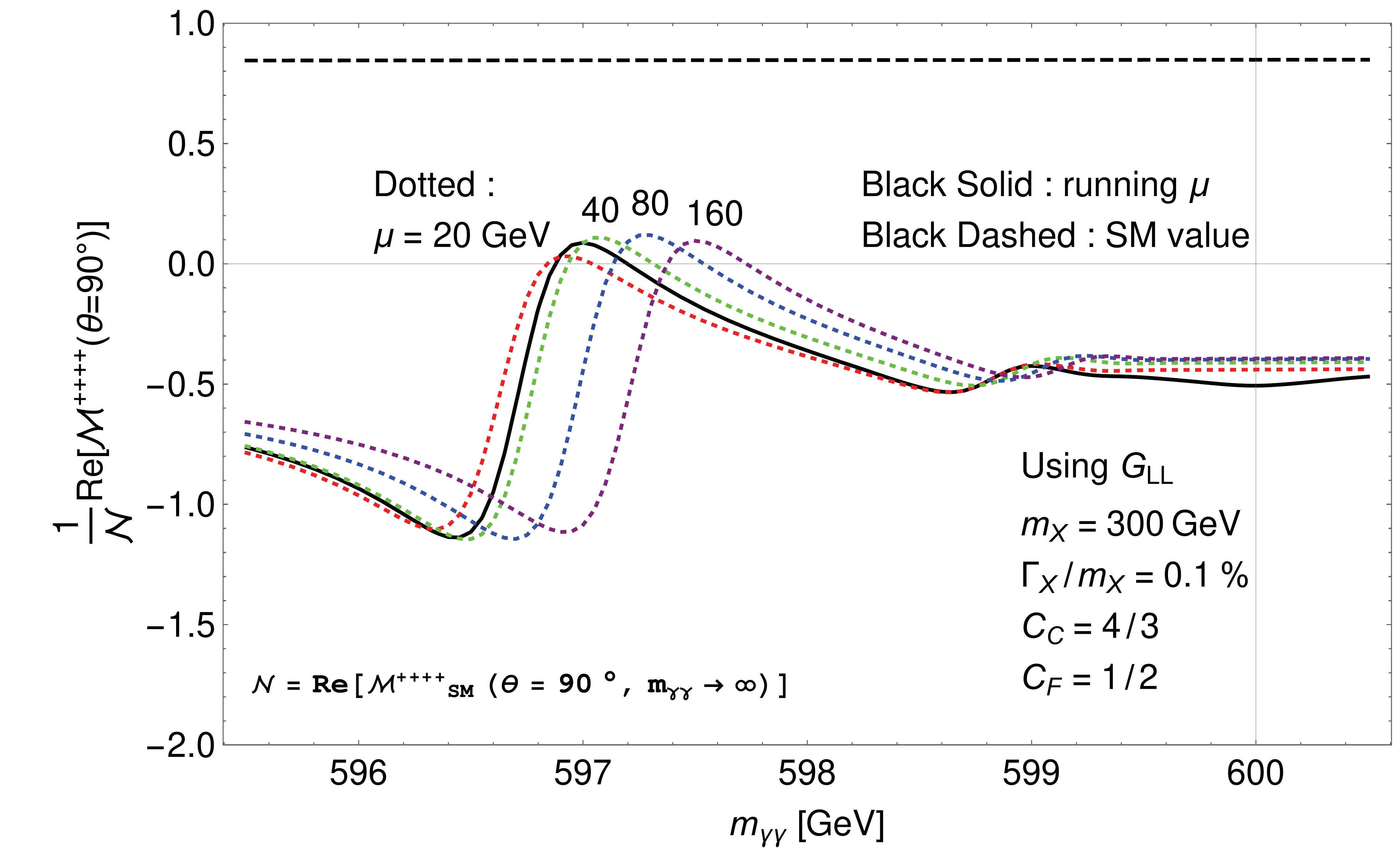}
\caption{}
\label{AmpNLOreal}
\end{subfigure}
\caption{ Normalized real part of amplitudes $gg\rightarrow \gamma\gamma$ (at an angle perpendicular to the beam line) using Eq. \eqref{WeUse} with the LO Green's function (a) and the LL Green's function (b). Normalization factor is chosen to be the real part of the standard model amplitude in large energy limit. Dotted lines are contributions from a fermion particle X with renormalization scales\red{, from left to right,} 20 GeV (Red), 40 GeV (Green), 80 GeV (Blue), and 160 GeV (Purple). Solid black line is using $\mu=\sqrt{m_X\sqrt{(m_{\gamma\gamma}-2m_X)^2+\Gamma_X^2}}$. Dashed black line is the contribution from the standard model quarks. Parameters for the fermion particle X are: $m_X=300$ GeV, $\Gamma_X/m_X=0.1 \%$, $C_C=4/3$, and $C_F=1/2$. Photon exchange ladder diagrams are neglected.}
\label{Ampreal}

\begin{subfigure}[b]{0.49\textwidth}
\includegraphics[width=\textwidth]{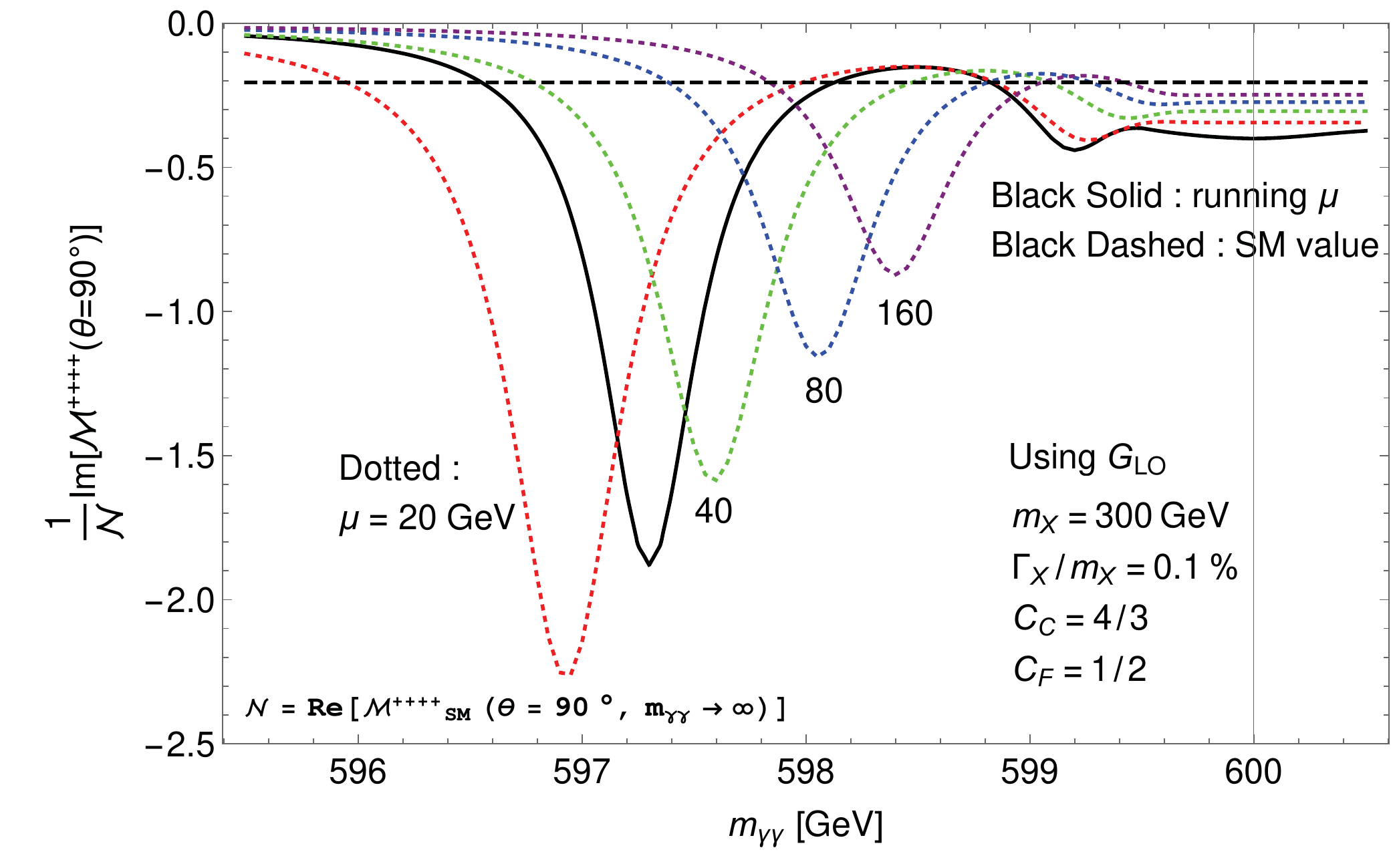}
\caption{}
\label{AmpLOim}
\end{subfigure}
\begin{subfigure}[b]{0.49\textwidth}
\includegraphics[width=\textwidth]{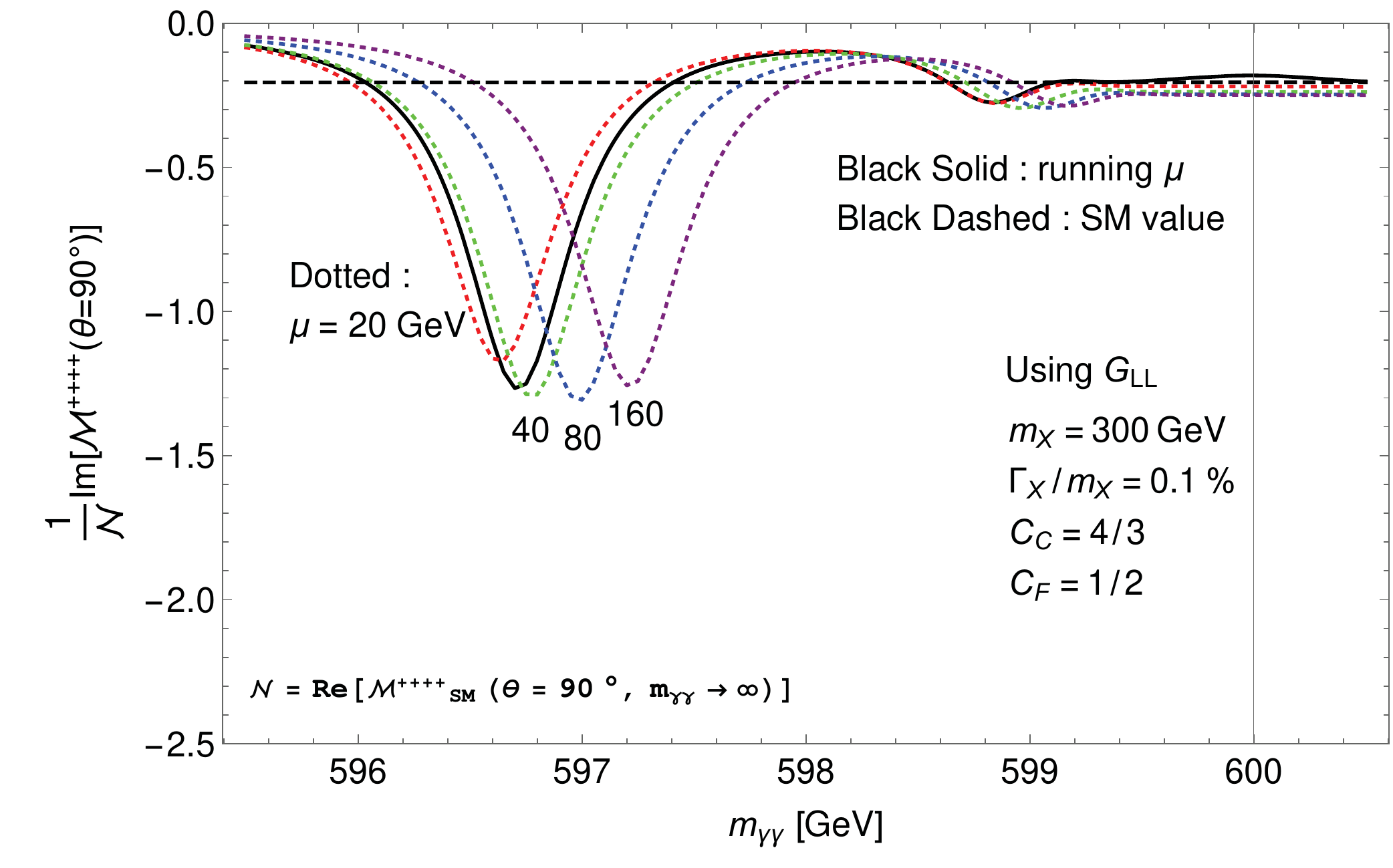}
\caption{}
\label{AmpNLOim}
\end{subfigure}
\caption{ Normalized imaginary part of amplitudes $gg\rightarrow \gamma\gamma$ (at an angle perpendicular to the beam line) using Eq. \eqref{WeUse} with the LO Green's function (a) and the LL Green's function (b). Normalization factor and parameters are the same as in Fig. \ref{Ampreal}.
}
\label{Ampim}
\end{figure}

In this section, we compare amplitude shapes using the leading order (LO) Green's function and the LL Green's function. Fig. \ref{Ampreal} and Fig. \ref{Ampim} describe real and imaginary parts of the amplitude shapes at the angle perpendicular to beam line as functions of invariant mass. Colored dotted lines are amplitude contributions only from particle X for renormalization scale\red{, from left to right,} 20 GeV (Red), 40 GeV (Green), 80 GeV (Blue), and 160 GeV (Purple). Black solid line corresponds to the choice of running $\mu$ in Eq. \eqref{runningmu}. Black dashed line is the amplitude involving standard model quarks. All of the amplitudes are normalized by the standard model amplitude at large energy limit. Because the top quark contribution is not saturated yet at 600 GeV, the dashed line in Fig. \ref{Ampreal} is slightly away from 1. 
Parameters for particle X are chosen to be: $m_X=300$ GeV, $\Gamma_X/m_X=0.1 \%$, $C_C=4/3$, and $C_F=1/2$. Photon exchange ladder diagrams are neglected.

 In QCD, a mass parameter that is free from renormalon is the 1S mass \cite{Hoang:1999zc}. This is where the 1S state resonance appears. The 1S mass is related to the parameter $m_X$ we use in the Schroedinger equation and the one-loop computation by
\bea
m_{\rm 1S}=m_X \left( 1- \f{1}{2}\f{C_C^2 \alpha_s^2}{4} \right)
\eea
for the LO Green's function, and by
\bea
m_{\rm 1S}=m_X \bigg( 1-  \f{1}{2}\f{C_C^2 \alpha_s^2}{4} \Big( 1  +\f{\alpha_s}{\pi} \big( \beta_0 \ln \left( \f{\mu e^{\gamma_E}}{m_X C_C \alpha_s} \right) +\f{a_1}{2} + \psi(2) \big) \Big) \bigg) 
\eea
for the LL Green's function \cite{Titard:1993nn}. Here, $\psi(x)=d \ln \Gamma(x)/ dx$ is the polygamma function. The $\beta_0$ and $a_1$ are defined below Eq. \eqref{nlopot}.

One can clearly see from the figures that using the LL Green's function decreases the renormalization scale dependence, since keeping the leading log lessens the $\mu$ dependence of the Coulomb potential. Thus, the LL Green's function is used in the following sections.

\section{Fermion Signal Shapes}
\label{fermionshape}

In this section, we show unpolarized cross section of $gg\rightarrow \gamma\gamma$ varying decay width $\Gamma_X$, color factor in Coulomb potential $C_C$, combined charge $C_X$, and electric charge $Q_X$ of a fermion particle X with $m_X$ fixed at 300 GeV. Coulomb potential from photon ladder exchanges is considered. Even if $Q$ is so large that $Q^2 \alpha$ is comparable to $\alpha_s$, its running effect is small and can be neglected. Gluon parton distribution function is considered using CTEQ6L data set \cite{Pumplin:2002vw}. The efficiency ($\sim 50\%$) of $P_T>0.4 m_{\gamma\gamma}$ cut and the K-factor ($\sim 150\%$) of gluon fusion production are not taken into account in our analysis. We assume that the K-factor does not change after the threshold resummation. In other words, we assume that most of the K-factor comes from other than the gluon ladder diagrams. Threshold resummation is a huge effect only around the threshold, while the K-factor affects the cross section at all energies.

Dependences on $\Gamma_X$, $C_C$, and $Q_X$ are shown in large $C_X$ limit. In this limit, the signal shape is $gg\rightarrow X\bar{X} \rightarrow \gamma \gamma$ without interference with standard model quark loops. The dependence on $C_X$ is shown for $gg \rightarrow \gamma\gamma$ process including interference with standard model quark loops. One should subtract the standard model part to see the signal shape.

\label{signalshapes}

 \subsection{Dependence on the width for large $C_X$}

\begin{figure}[t]
\includegraphics[width=0.6\textwidth]{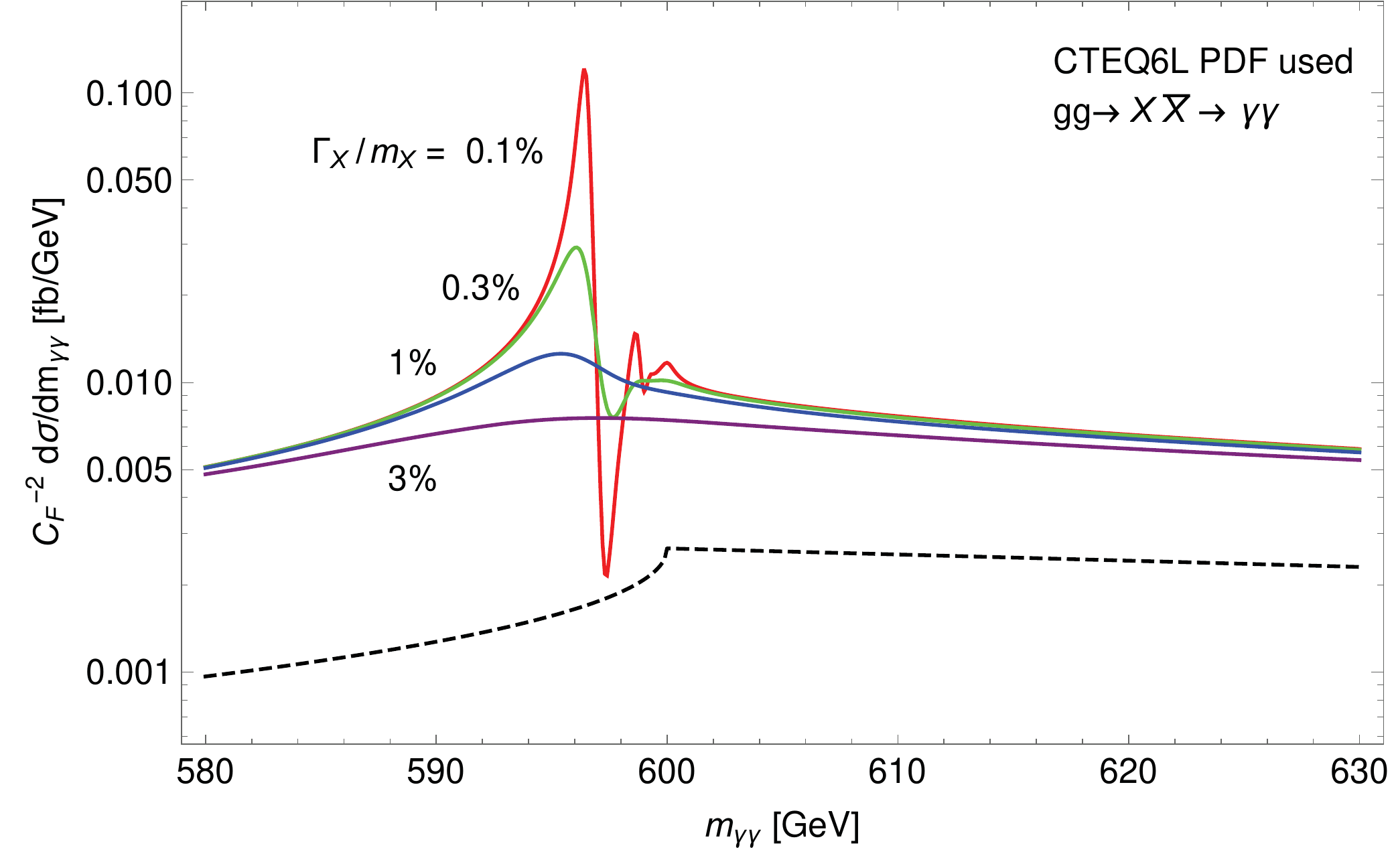}
\caption{ Scattering cross section of $gg \rightarrow \gamma\gamma$ as a function of invariant mass, mediated by a fermion with: $m_X = 300$ GeV, $Q_X=1$, and $C_C$=4/3. Only the particle X contribution is shown with normalization of $C_X^{-2}$. Running of strong couplings, $\alpha_s(m_{\gamma\gamma})$ for overall factor and $\alpha_s(\sqrt{m_X\sqrt{(m_{\gamma\gamma}-2m_X)^2+\Gamma_X^2}})$ for Coulomb potential is considered. QED couplings, $\alpha=1/127$ for overall factor and $\alpha=1/130$ for Coulomb potential are used. 
 $\Gamma_X/m_X$ is 0.1\% (Red), 0.3\% (Green), 1\% (Blue), and 3\% (Purple). Black dashed line represents the one-loop result.  }
\label{plot1}
\end{figure}

As decay width decreases, signal shape becomes sharper and higher. At the same time, area under the curve increases. Actually, for very small $\Gamma_X$, narrow width approximation can be applied, to each bound state excitation, to separate production and decay parts. However, for large decay width, it is important to also keep real parts of amplitudes including the effective $gg\gamma\gamma$ vertex. Fig. \ref{plot1} shows signal shapes, $gg\rightarrow X\bar{X} \rightarrow \gamma \gamma$ for various decay widths of the fermion particle X with: $m_X = 300$ GeV, $Q_X=1$, and $C_C$=4/3.
$\Gamma_X/m_X$ is 0.1\% (Red), 0.3\% (Green), 1\% (Blue), and 3\% (Purple). Black dashed line represents the one-loop result. Running of strong couplings, $\alpha_s(m_{\gamma\gamma})$ for overall factor and $\alpha_s(\sqrt{m_X\sqrt{(m_{\gamma\gamma}-2m_X)^2+\Gamma_X^2}})$ for Coulomb potential is considered. QED couplings, $\alpha=1/127$ for overall factor and $\alpha=1/130$ for Coulomb potential are used. QED coupling is about $1/127$ for the scale of order $100$ GeV and about $1/130$ for the scale of order $10$ GeV which is typical momentum scale exchanged in the ladder diagrams.

 \subsection{Dependence on the color factor for large $C_X$}
\label{wierdsommer}

If the quadratic Casimir of the particle X or that of the bound state changes, then the color factor for Coulomb potential $C_C$ also changes as $C_C = C_{2,X}-\f{1}{2}C_{2,X\bar{X}}$. For diphoton process, $C_{2,X\bar{X}}$ is always $0$ because diphoton is color singlet. On the other hand, the $gg \rightarrow g \gamma$ process should carry $C_{2,X\bar{X}}=3$. This gives $C_C=-1/6$ for the particle X in the fundamental representation. In $gg \rightarrow gg$ process, we can obtain $C_C=-1$ for the particle in the octet representation and the bound state in the representation of dimension $\textbf{27}$.

\begin{figure}[t]
\includegraphics[width=0.6\textwidth]{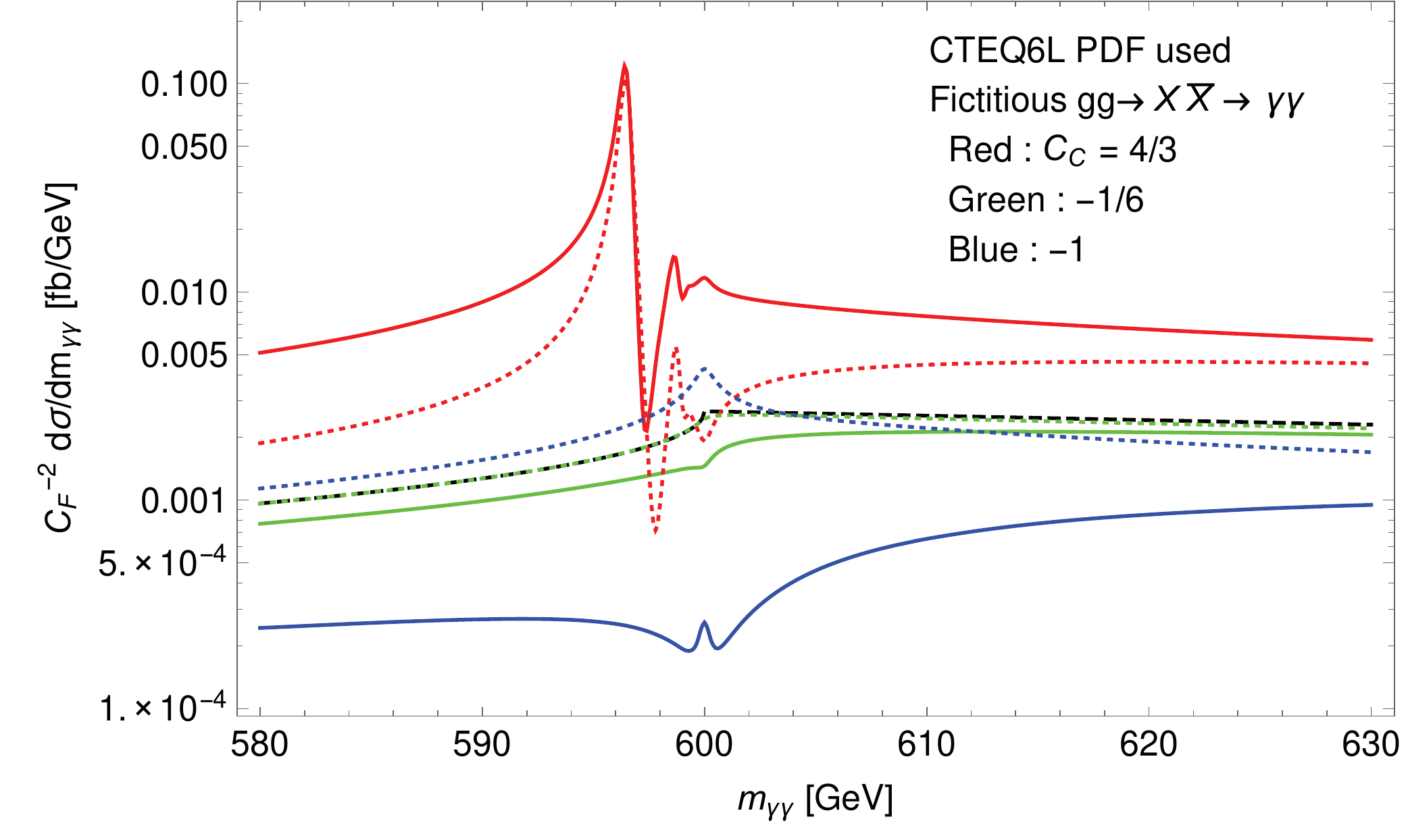}
\caption{ Fictitious $gg \rightarrow g\gamma$ and $gg \rightarrow gg$ with no gluon self-interaction (i.e., box loop only). Purpose of this is to show strange behavior coming from $\ln (m\alpha_s/m\beta)$ after summation over excitations for small velocity if we do not properly consider the running of effective tree level vertex. The large $C_X$ limit is shown with normalization of $C_X^{-2}$.
$C_C$ is $4/3$ (Red), $-1/6$ (Green), and $-1$ (Blue). Black dashed line represents the one-loop result. Solid lines are using Eq. \eqref{WeUse}. Dotted lines are without the third term (the log term) in Eq. \eqref{WeUse}. Other parameters are as in Fig. \ref{plot1} with $\Gamma_X/m_X=0.1\%$.
 }
\label{plot3}
\end{figure}

Fig. \ref{plot3} shows "fictitious" $gg \rightarrow X\bar{X} \rightarrow \gamma\gamma$ process merely changing $C_C$ to be 4/3 (Red), -1/6 (Green), and -1 (Blue). The fictitious cross section with $C_C=-1/6$ is proportional to the cross section for $g \gamma$ process. The fictitious cross section with $C_C=-1$ is proportional to the partial cross section for $gg$ process; it is only partial because four gluon vertex in full theory is not considered. Solid lines are using the correct form, Eq. \eqref{WeUse} with the LL Green's function and dotted lines are, again, using Eq. \eqref{WeUse} with the LL Green's function but without the third term (log term) in Eq. \eqref{WeUse}.
Lose of the log term means that the running of $gg\gamma\gamma$ vertex is not considered properly. Dashed black line is one-loop cross section. We can see from the solid lines that, if the running is properly considered, then Sommerfeld suppression is obtained for negative $C_C$ or repulsive potential. The blue dotted line shows that ignoring the log term would result in Sommerfeld "enhancement" for repulsive potential.

The unphysical enhancement is due to the choice of $\mu \sim m \beta$ and the excitation summation in the Green's function, Eq. \eqref{green}. Choosing $\mu \sim m \beta$, the log term in the Green's function is gone. If the running of effective tree level vertex is taken into account, then we do not loose the log because $A(m_X \beta)=A(m_X)+B C_C \alpha_s \ln \left( \f{m_X \beta}{m_X}\right)$. Ignoring the running means that we are setting $A(m_X \beta)=A(m_X)$ and we do not see the explicit $\ln (m_X \beta)$ anymore.

Other than the explicit log term, $C_C \alpha_s (\mu) \ln (\sqrt{-m_X E}/\mu)$, there is a hidden log term in the excitation summation which is the third term of Eq. \eqref{green}. 
When $E/E_1$ is much smaller than 1, there exists $n$ such that up to $n$ we can ignore $E$. 
Then, the summation is $\Sigma_n E_n/\sqrt{E_n} \propto \Sigma_n 1/n$ which is approximated by log function:
\bea
  &&-\frac{2}{\sqrt{m_X}}\sum_{n=1}^{\infty}\frac{E_n}{\sqrt{(-E-i \epsilon)}-\text{sign}(C_C)\sqrt{E_n} }   \label{implicitlog}   \\
  &&= C_C \alpha_s \left(  \gamma+ \psi \left(1-\f{C\alpha_s}{2 \sqrt{-E/m}}\right) \right) 
   \simeq C_C \alpha_s \left( \gamma + \ln \left(\f{-C_C \alpha_s/2}{\sqrt{-E/m}}\right)\right)-\sqrt{\f{-E}{m}}+\mathcal{O} \left(\left(\sqrt{\f{-E}{m}}\right)^2\right), \nn
\eea
unless $E=E_n$ for some $n$. If the running was properly considered, there would have been the explicit $\ln (\sqrt{-m_X E})$ term that cancels the same log term in Eq. \eqref{implicitlog}. It is this log term that gives the strange behavior of the blue dotted line in Fig. \ref{plot3} for small $\sqrt{-m_X E}$. In order to illustrate this point, $\Gamma_X/m_X$ is chosen to be $0.1\%$. Parameters except $C_C$ and $\Gamma_X/m_X$ are kept to be the same as in Fig. \ref{plot1}.

\begin{figure}[t]
\includegraphics[width=0.6\textwidth]{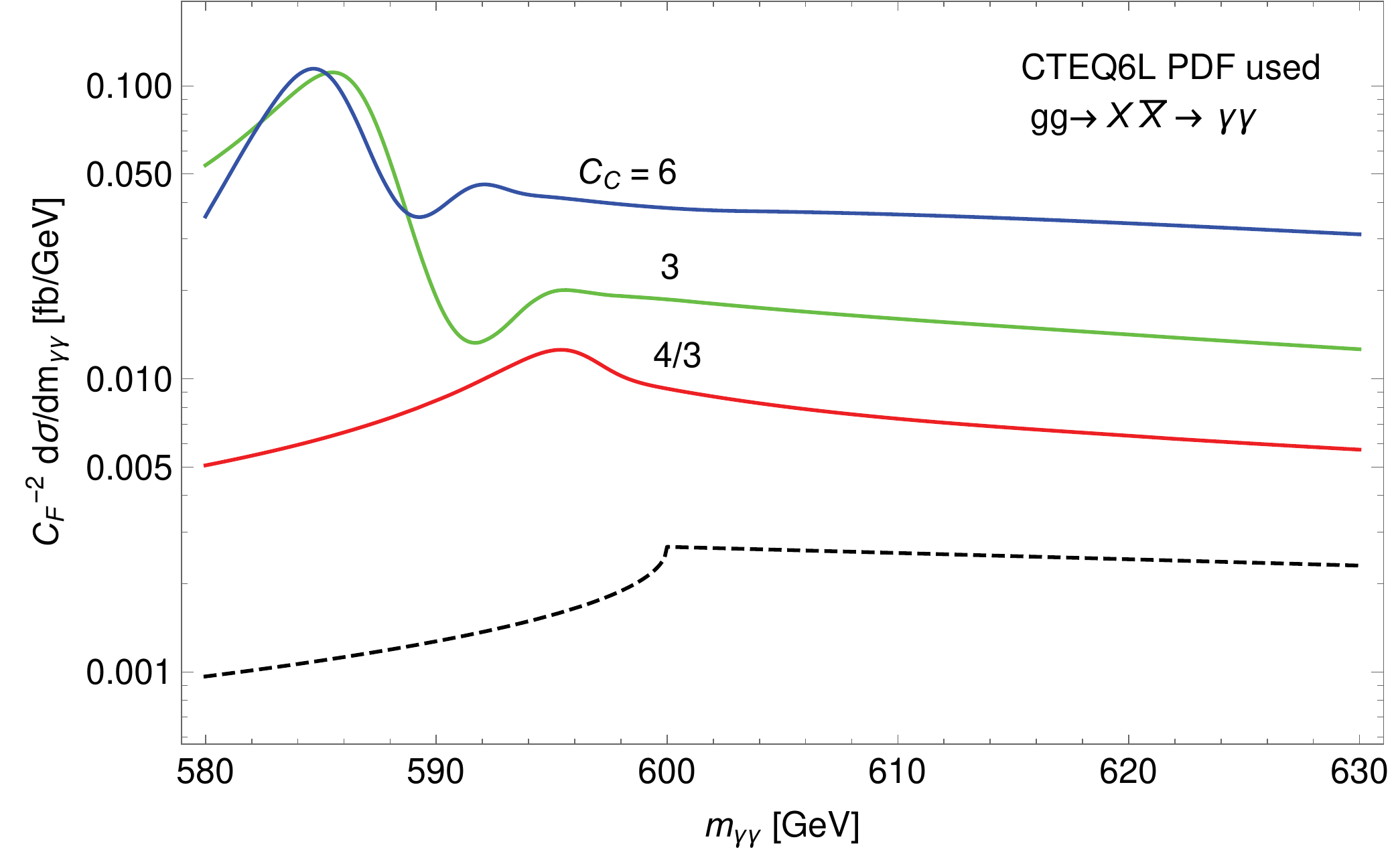}
\caption{ Scattering cross section of $gg \rightarrow \gamma\gamma$ as a function of invariant mass. Only the particle X contribution is shown with normalization of $C_X^{-2}$. $C_C$ is $4/3$ (Red), 3 (Green), and 6 (Blue). Black dashed line represents the one-loop result.
Other parameters are as in Fig. \ref{plot1} with $\Gamma_X/m_X=1\%$.
}
\label{plot2}
\end{figure}

For $gg\rightarrow \gamma\gamma$ process, different choices of $C_C$ can come from choosing different particle representation under $SU(3)_C$. However, next the smallest possible $C_C$ is 3 (octet particle X) and this gives already too large $C_C \alpha_s$ which is one of the expansion parameters.  The signal shape is shown Fig. \ref{plot2} for $C_C$=3 or 6. Although it is perturbatively meaningless, we guess that the shown tendency of getting larger signal for larger $C_C$ is true.

 \subsection{Dependence on $C_X$}

\begin{figure}[t]
\begin{subfigure}[b]{0.49\textwidth}
\includegraphics[width=\textwidth]{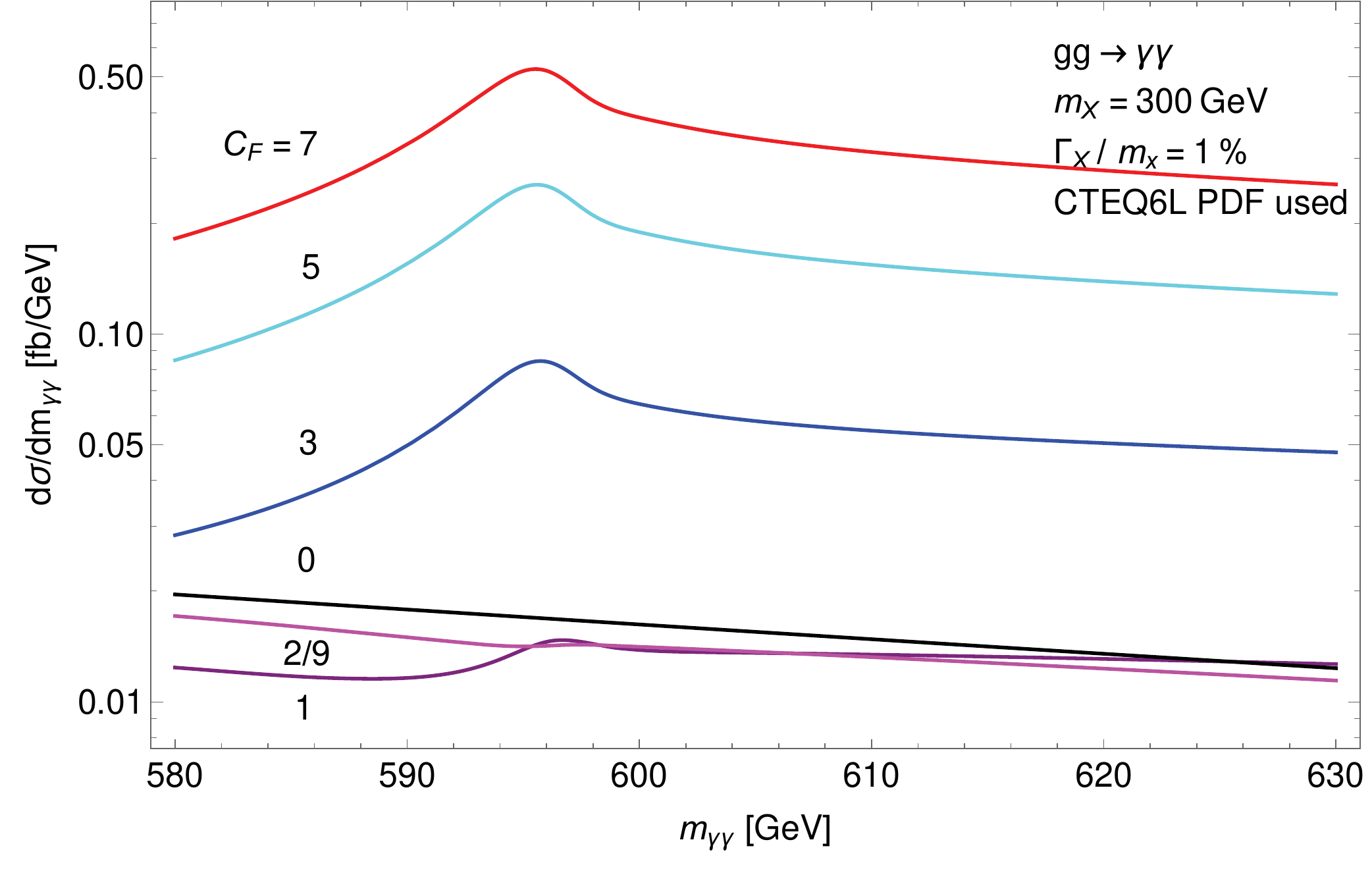}
\caption{}
\label{plot4out}
\end{subfigure}
\begin{subfigure}[b]{0.49\textwidth}
\includegraphics[width=\textwidth]{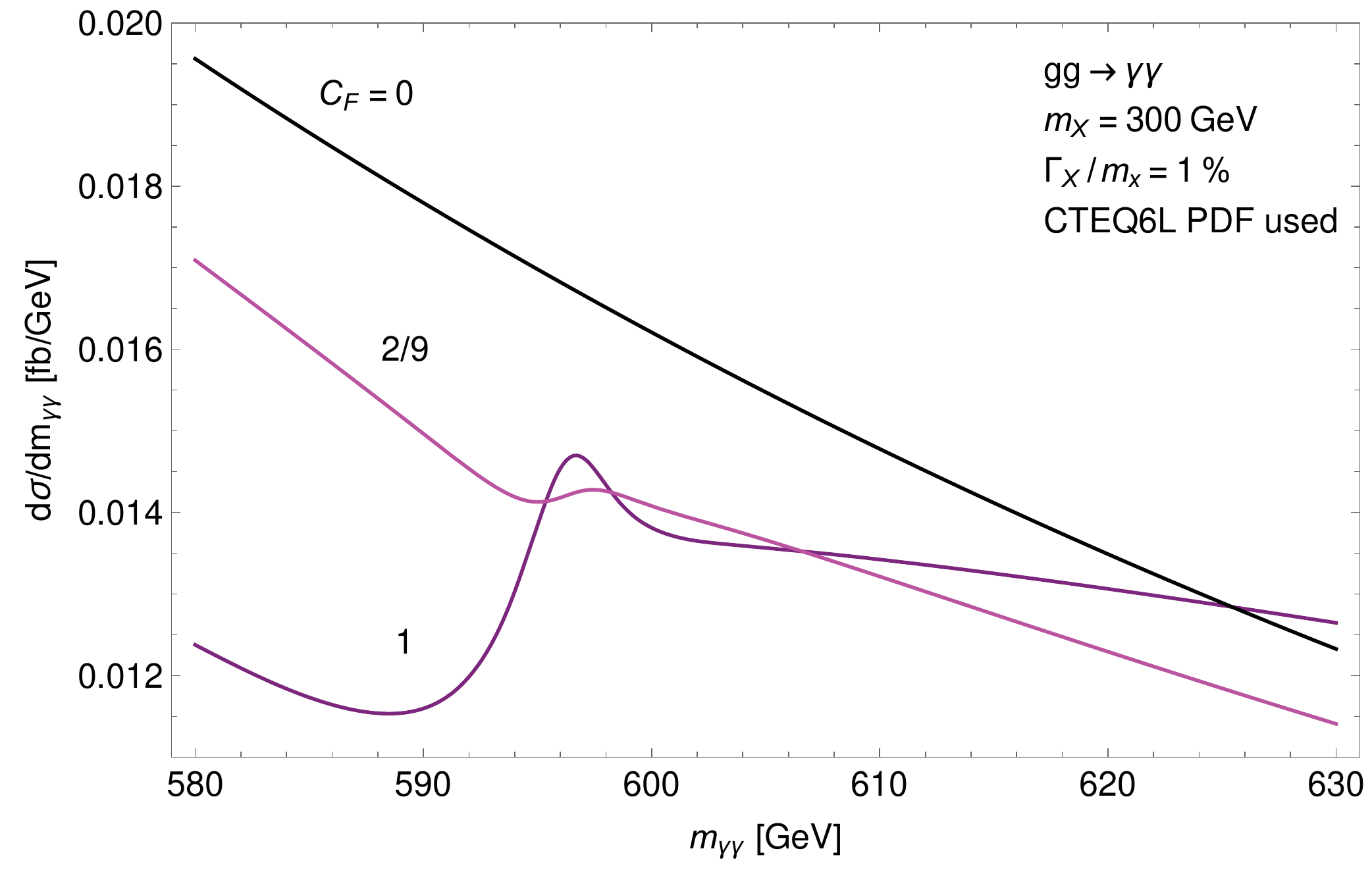}
\caption{}
\label{plot4in}
\end{subfigure}
\caption{ Scattering cross section of $gg \rightarrow \gamma\gamma$ through standard model quarks and the fermion particle X as a function of invariant mass (a) and magnified in (b).
$C_X$ is 7 (Red), 5 (Cyan), 3 (Blue), 1 (Purple), and $2/9$ (Magenta). Black line represents standard model result, $C_X=0$.
Other parameters are as in Fig. \ref{plot1} with $\Gamma_X/m_X=1\%$.
}
\label{plot4}
\end{figure}

The previous sections focused on large $C_X$ limit and interference effect was neglected. For small $C_X$, interference between X particle loop and  standard model quark loops is important. Fig. \ref{plot4} shows the cross section of $gg \rightarrow \gamma\gamma$ through standard model quarks and the fermion particle X as a function of diphoton invariant mass.
$C_X$ is 7 (Red), 5 (Cyan), 3 (Blue), 1 (Purple), and $2/9$ (Magenta). Black line represents standard model result, $C_X=0$.
Other parameters are as in Fig. \ref{plot1} with $\Gamma_X/m_X=1\%$. Fig. \ref{plot4in} is a magnified version of Fig. \ref{plot4out}.

What we usually call signal shape will be obtained by adding this $gg\rightarrow\gamma\gamma$ to other background process like $q\bar{q}\rightarrow\gamma\gamma$ and then subtracting the standard model fitting function. Roughly speaking, it would look like colored lines minus black line in Fig. \ref{plot4}. The magenta line has the same $C_C=2/9$ as top quark does. Top quark contribution after resummation is shown in Ref. \cite{Chway:2015lzg}.

 \subsection{Dependence on $Q_X$ for large $C_X$}

\begin{figure}[t]
\begin{subfigure}[b]{0.49\textwidth}
\includegraphics[width=1.0\textwidth]{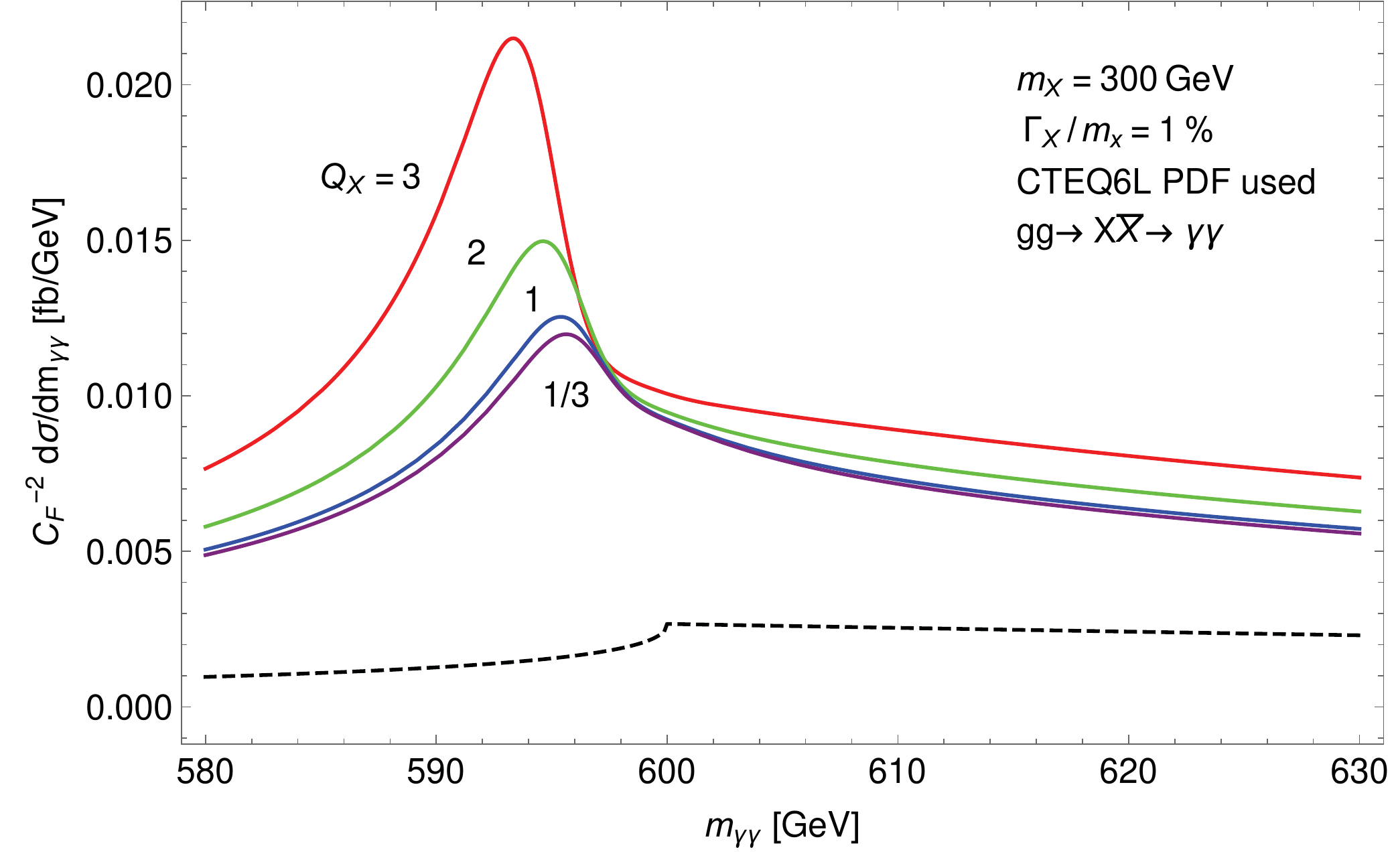}
\caption{}
\label{plot5}
\end{subfigure}
\begin{subfigure}[b]{0.49\textwidth}
\includegraphics[width=1.0\textwidth]{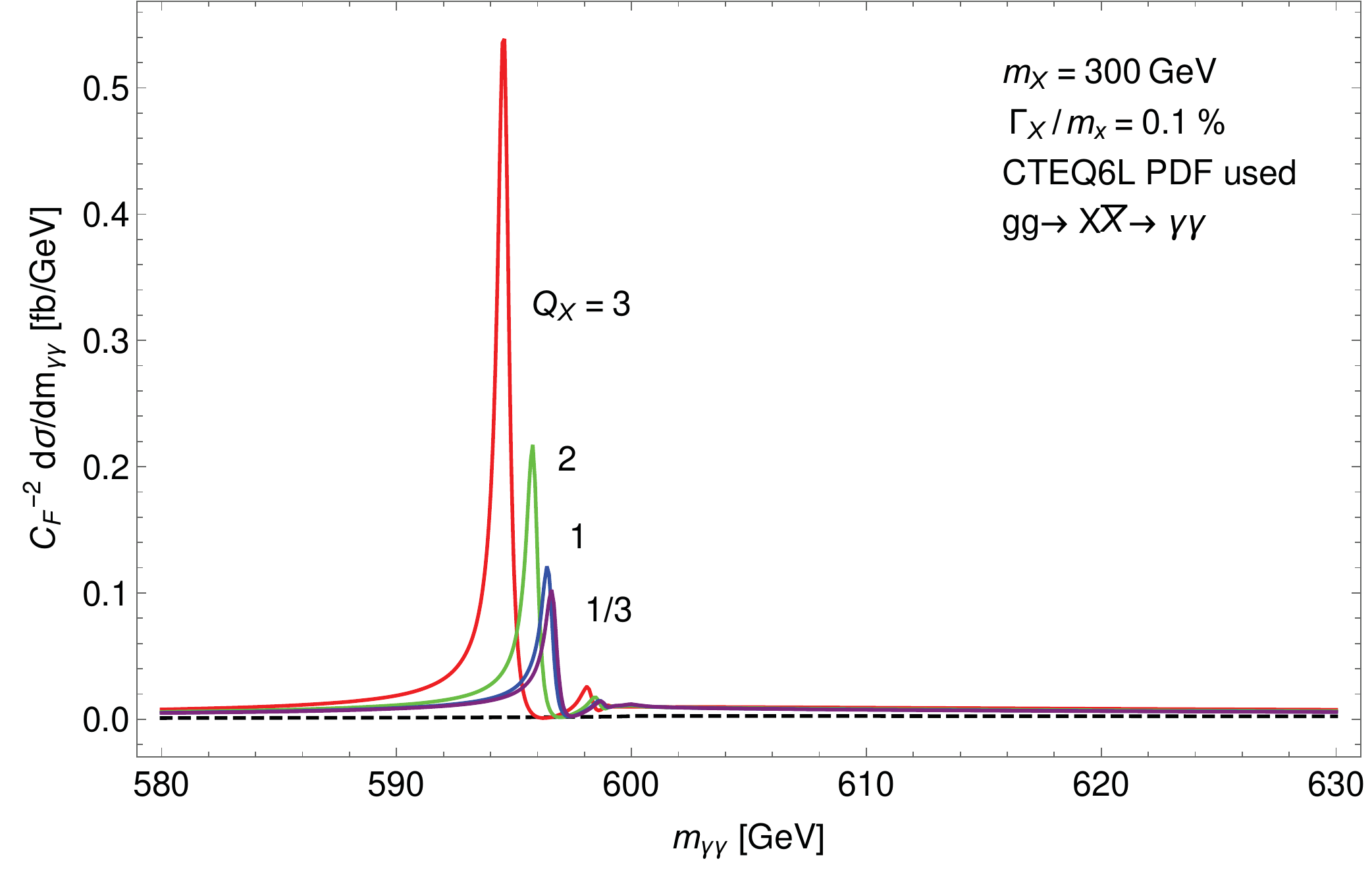}
\caption{}
\label{plot6}
\end{subfigure}
\caption{Scattering cross section normalized with $C_X^{-2}$ of $gg \rightarrow \gamma\gamma$ only through the particle X as a function of invariant mass.
$Q_C$ is 3 (Red), 2 (Green), 1 (Blue), and $1/3$ (Purple). Black dashed line represents the one-loop result.
Other parameters are as in Fig. \ref{plot1} with $\Gamma_X/m_X$=1\% (a) and 0.1\% (b).
}
\end{figure}

There are ladder diagrams of not only the gluons, but also photons. \red{The summation of the photon ladder diagrams is combined with that of the gluon ladders by replacing the coefficient of the Coulomb potential $C_C\alpha_s$ by $C_C\alpha_s+Q_X^2 \alpha$ in the Schroedinger equation Eq. \eqref{schroedinger}.} Dependence on electric charge of particle X is shown in Fig. \ref{plot5} for $\Gamma_X/m_X=1\%$ and Fig. \ref{plot6} for $\Gamma_X/m_X=0.1\%$. Running of strong couplings, $\alpha_s(m_{\gamma\gamma})$ for overall factor and $\alpha_s(\sqrt{m_X\sqrt{(m_{\gamma\gamma}-2m_X)^2+\Gamma_X^2}})$ for Coulomb potential is considered. Overall electric couplings, $\alpha=1/127$ and Coulomb electric coupling, $\alpha=1/130$ are fixed without running. QED coupling is about $1/127$ for the scale of order $100$ GeV and about $1/130$ for the scale of order $10$ GeV which is typical momentum scale exchanged in the ladder diagrams. Electric charge $Q_X$ is 3 (Red), 2 (Green), 1 (Blue), and $1/3$ (Purple). Black dashed line represents the one-loop result.

Larger charge gives larger cross section. Considering photon ladder resummation is more important in case the particle X has smaller decay width. 

\section{Scalar Signal Shapes}
\label{signalshapesscalar}

\begin{figure}[t]
\includegraphics[width=0.6\textwidth]{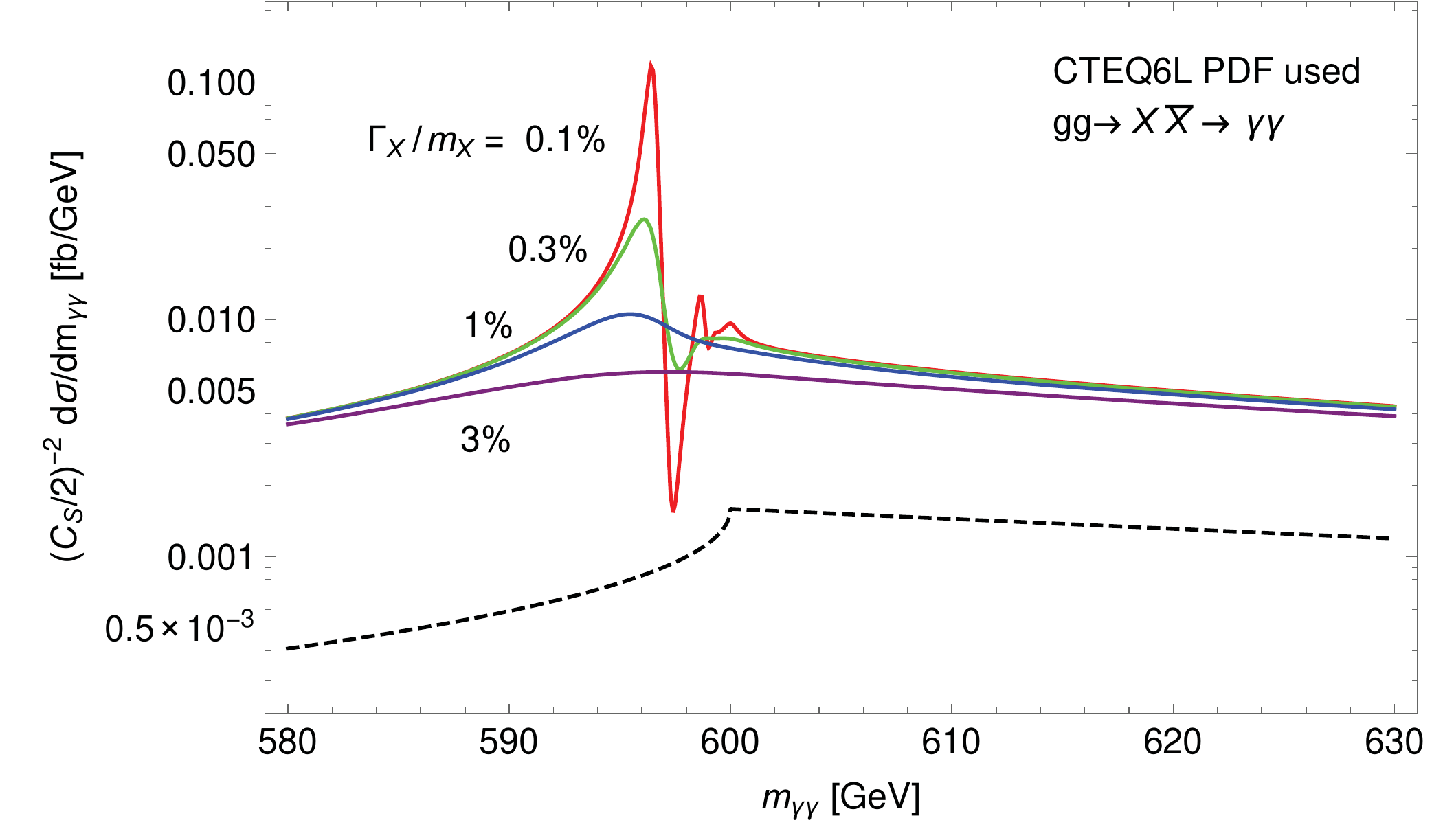}
\caption{ Scattering cross section of $gg \rightarrow \gamma\gamma$, mediated by a scalar X, as a function of invariant mass. 
 $\Gamma_X/m_X$ is 0.1\% (Red), 0.3\% (Cyan), 1\%(Blue), and 3\%(Purple). Black dashed line represents the one-loop result. Parameters are the same as in Fig. \ref{plot1}.
 }
\label{plot1s}
\begin{subfigure}[b]{0.49\textwidth}
\includegraphics[width=\textwidth]{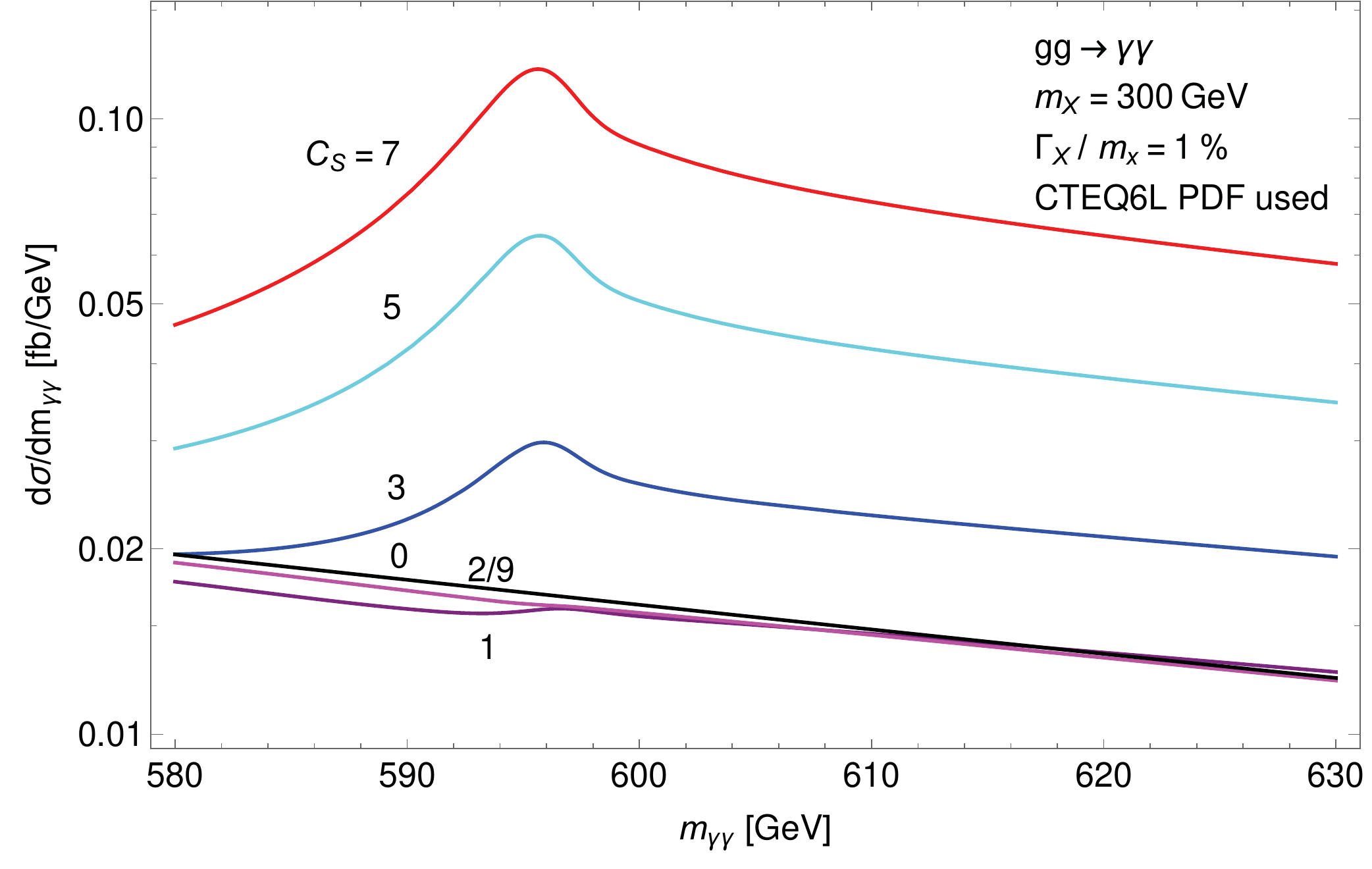}
\caption{}
\label{plot4sout}
\end{subfigure}
\begin{subfigure}[b]{0.49\textwidth}
\includegraphics[width=\textwidth]{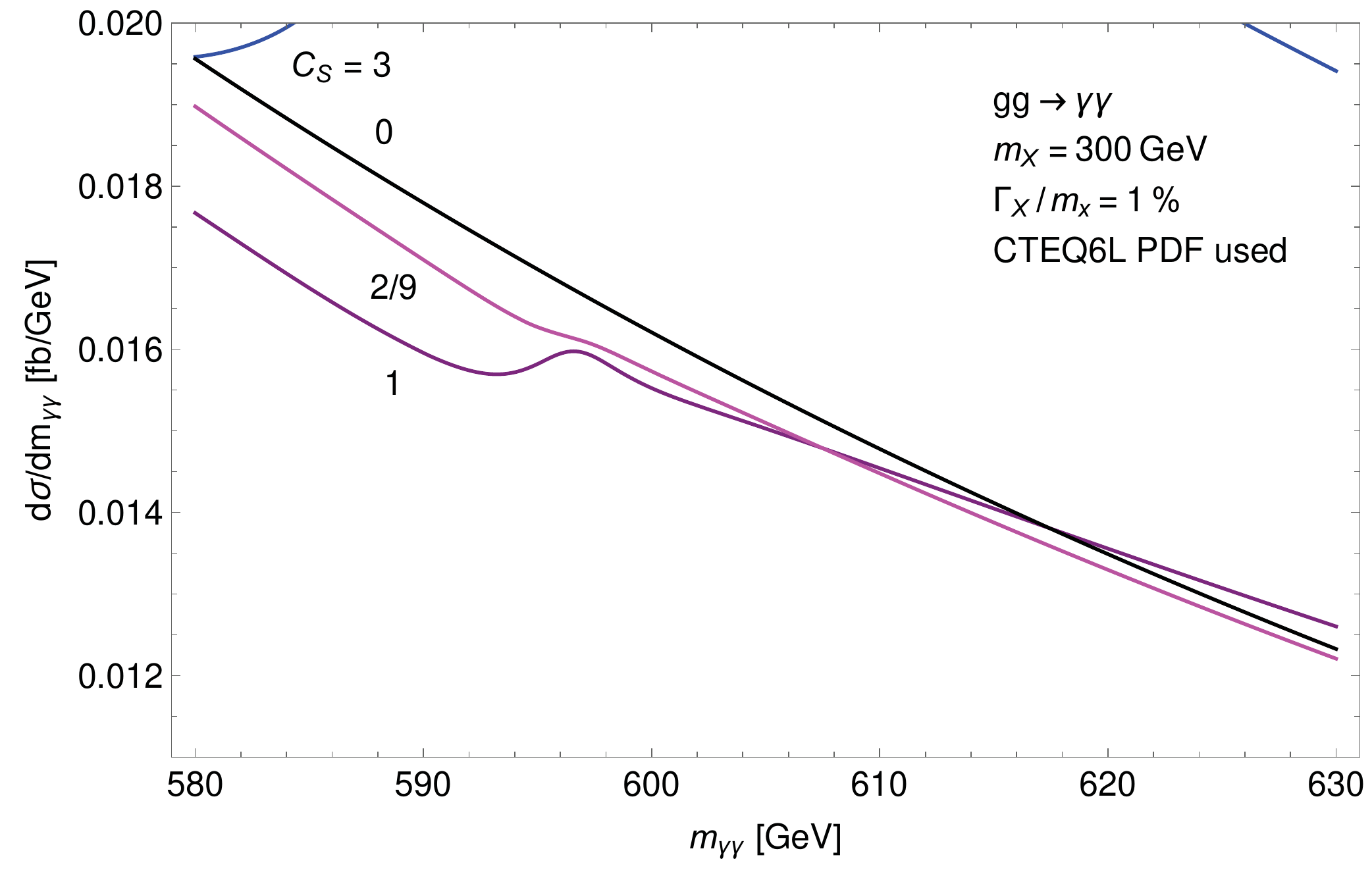}
\caption{}
\label{plot4sin}
\end{subfigure}
\caption{ Scattering cross section of $gg \rightarrow \gamma\gamma$, through standard model quarks and the scalar X, as a function of invariant mass. 
The $C_X$ is 7 (Red), 5 (Cyan), 3 (Blue), 1 (Purple), and $2/9$ (Magenta). Black line represents standard model result, $C_C=0$.
Other parameters are as in Fig. \ref{plot1} with $\Gamma_X/m_X=1\%$.
}
\label{plot4s}
\end{figure}

\begin{figure}[th]
\begin{subfigure}[b]{0.49\textwidth}
\includegraphics[width=1.0\textwidth]{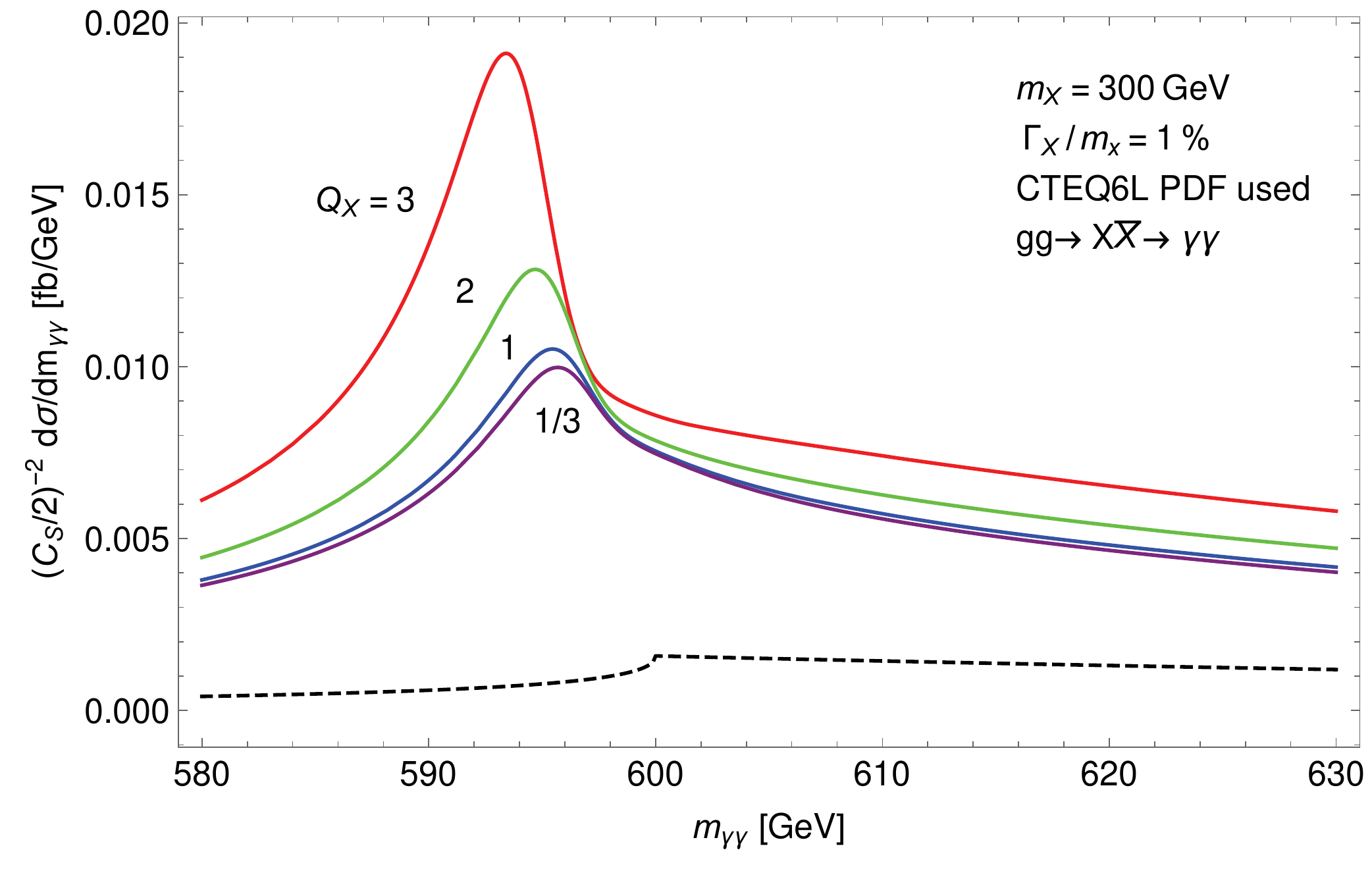}
\caption{}
\label{plot5s}
\end{subfigure}
\begin{subfigure}[b]{0.49\textwidth}
\includegraphics[width=1.0\textwidth]{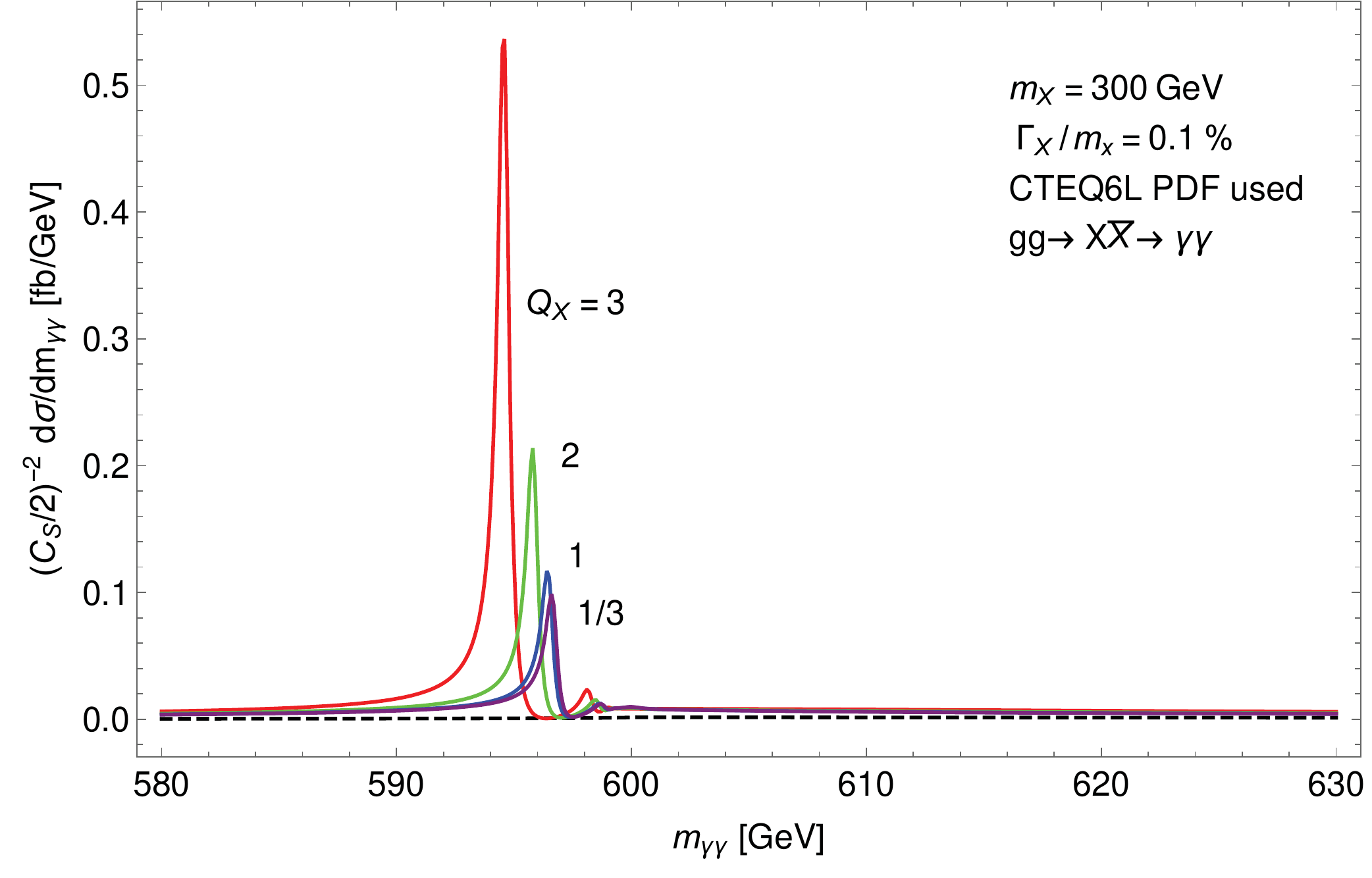}
\caption{}
\label{plot6s}
\end{subfigure}
\caption{ Scattering cross section, mediated by the scalar X, normalized with $C_X^{-2}$, of $gg \rightarrow \gamma\gamma$ as a function of invariant mass. 
$Q_X$ is 3 (Red), 2 (Green), 1 (Blue), and $1/3$ (Purple). Black dashed line represents the one-loop result.
Other parameters are as in Fig. \ref{plot1} with $m_X=300$ GeV and $\Gamma_X/m_X$=1\% (a) and 0.1\% (b).
 }
\label{plot56s}
\end{figure}

In this section we provide corresponding signal shapes in case the particle X is a scalar. 
The main difference from the fermion case originates from the fact that the ratio of  $\mathcal{M}_{\rm UV 1-loop} (2m_X)$ to $B$ in Eq. \eqref{AmG} is different.
For large $C_X$, scattering cross section of $gg \rightarrow \gamma\gamma$, mediated by a scalar X, is shown as a function of invariant mass in Fig. \ref{plot1s}. 
Four different choices of the widths are taken; $\Gamma_X/m_X$ is 0.1\% (Red), 0.3\% (Cyan), 1\%(Blue), and 3\%(Purple). Black dashed line represents the one-loop result. All the arameters used in Fig. \ref{plot1s} are the same as in Fig. \ref{plot1}

For moderate or small $C_X$, we include interference in Fig. \ref{plot4s}.
Scattering cross section of $gg \rightarrow \gamma\gamma$, through standard model quarks and the scalar X, is shown as a function of invariant mass. 
The $C_X$ is 7 (Red), 5 (Cyan), 3 (Blue), 1 (Purple), and $2/9$ (Magenta). Black line represents standard model result, $C_C=0$.
Other parameters are as in Fig. \ref{plot1} with $\Gamma_X/m_X=1\%$.

Finally, we show dependence on the electric charge for large $C_X$ in Fig. \ref{plot56s}.
Scattering cross section, mediated by the scalar X, normalized with $C_X^{-2}$, of $gg \rightarrow \gamma\gamma$ is shown as a function of invariant mass. 
Four difference choices of the charges are taken; $Q_X$ is 3 (Red), 2 (Green), 1 (Blue), and $1/3$ (Purple). Black dashed line represents the one-loop result.
Other parameters are as in Fig. \ref{plot1} with $m_X=300$ GeV and $\Gamma_X/m_X$=1\% (a) and 0.1\% (b).

\section{Exclusion Plots}
\label{exclusionsection}

In order to make exclusion plots, we assume $pp\to\gamma\gamma$ differential cross section can be separated into two parts, gluon initiated process and the others. For non-gluon initiated process, we assume it can be fitted by a smooth function, 
{\small
\bea
\frac{d}{dm_{\gamma\gamma}}\left(\sigma(pp\to\gamma\gamma)-\sigma(gg\to\gamma\gamma)\right)=N (1-x^{1/3})^b x^{a_0},
\eea}\noindent 
where $x=m_{\gamma\gamma}/\sqrt{S}$ for the center of mass energy $\sqrt{S}$ and $N$ is normalization factor which depends on two fitting parameters, $a_0$ and $b$ \cite{ATLAS:2016eeo, Aaltonen:2008dn, Aad:2014eha}.
\red{This assumption was validated in Ref. \cite{ATLAS:2016eeo}. Unlike the references where the background function is fitted for $pp\rightarrow \gamma\gamma$ process, we further assume that the background function well describes non-gluon initiated process alone too, of course with different values of $a_0$ and $b$ than in the references.}
For gluon initiated process, we follow \red{the} matching procedure and \red{the} resummation method that we described in \red{the} previous sections. Here, we use LL QCD potential and we choose RG scales\red{: overall coupling scale $\mu_{\rm hard}=m_{\gamma\gamma}$, QCD factorization scale $\mu_{\rm factorization}=m_{\gamma\gamma}$ and the ladder exchange scale $\mu_{\rm soft}=m_X^{1/2}\left((m_{\gamma\gamma}-2m_X)^2+\Gamma_X^2\right)^{1/4}$}. As was discussed in section \ref{fermionshape}, K-factor and the cut selection efficiency were not considered.

Unlike in Ref.\cite{Chway:2015lzg}, we use maximum binned likelihood estimation. Null hypothesis corresponds to using standard model gluon initiated cross section while signal hypothesis is that of standard model plus new particle $X$. \red{The procedure to obtain exclusion plots is described in detail in Appendix \ref{LLapen}.} For current exclusion plots, we use recent ATLAS $15.4 {\rm fb}^{-1}$ data \cite{ATLAS:2016eeo} and for expected exclusion plots, we assume that the best fitted values of the parameters $a_0$ and $b$ for the current data are the true values.

\begin{figure}[t]
\begin{subfigure}[b]{0.49\textwidth}
\includegraphics[width=1.0\textwidth]{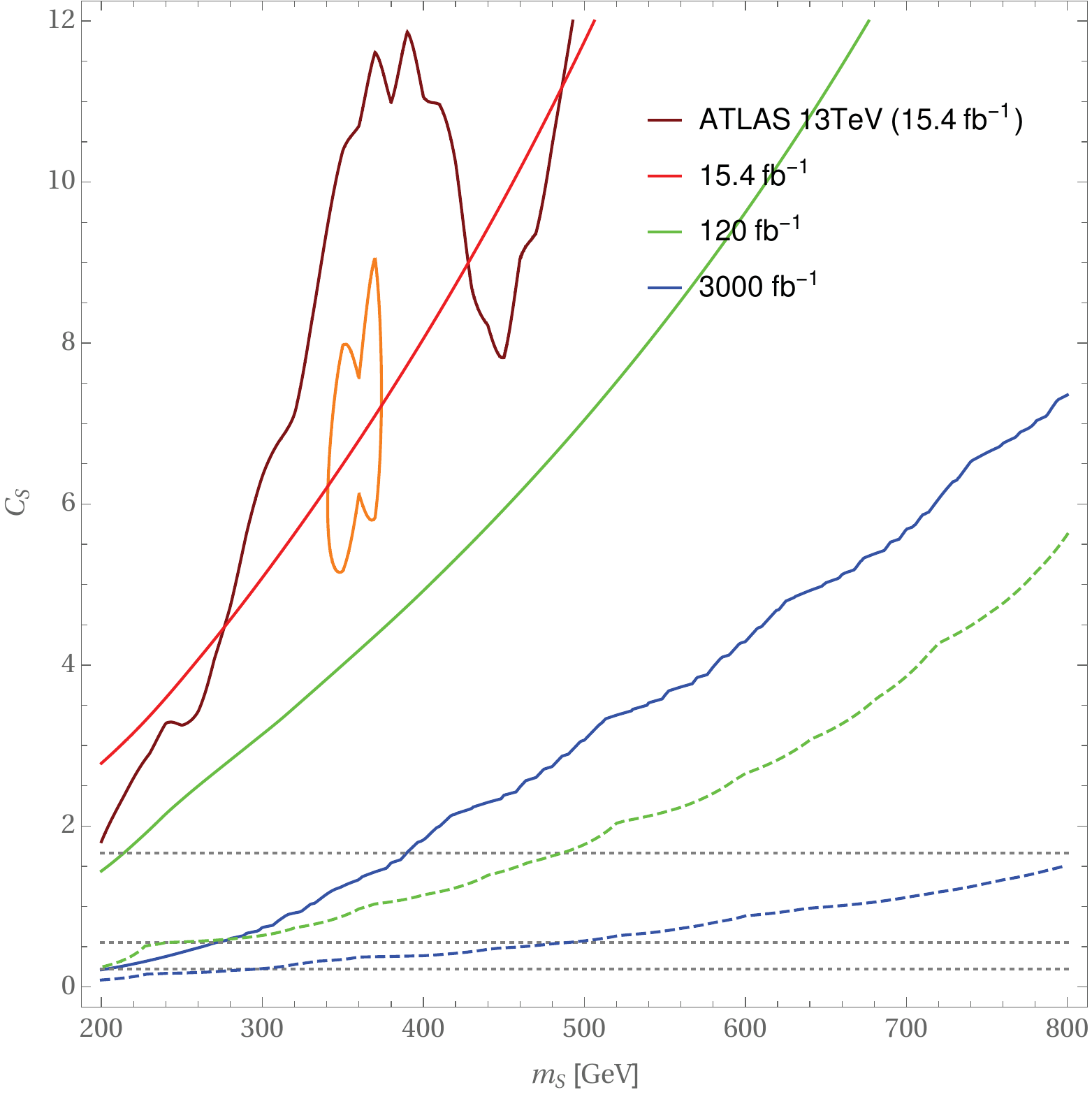}
\caption{scalar}
\label{contours}
\end{subfigure}
\begin{subfigure}[b]{0.49\textwidth}
\includegraphics[width=1.0\textwidth]{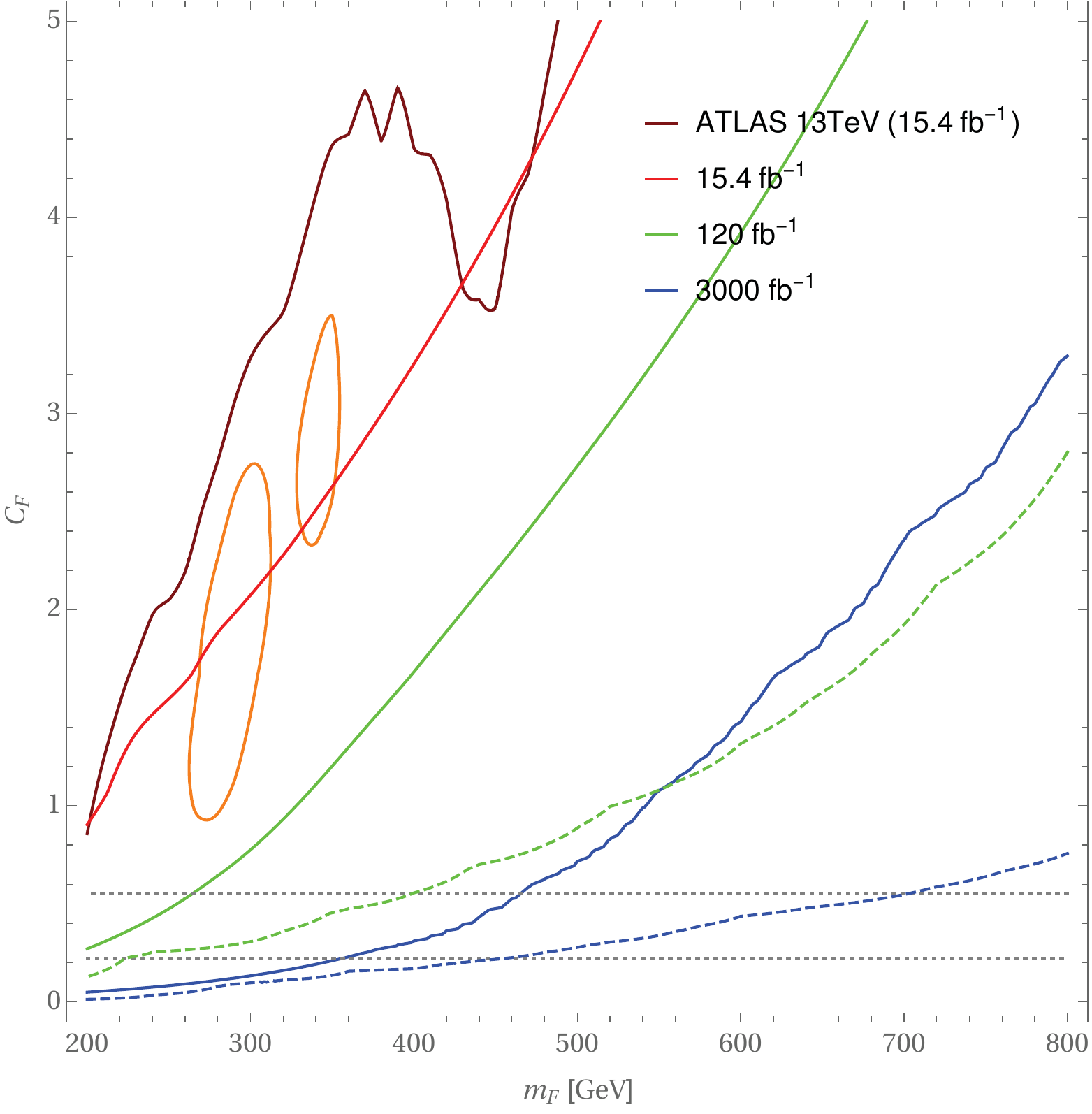}
\caption{fermion}
\label{contourf}
\end{subfigure}
\caption{95\% C.L. exclusion limits on $m_X$ and $C_X$ parameter space. Dark red curve corresponds to the current exclusion plot and red, green and blue curves correspond to the expected exclusion curves with integrated luminosity, 15.4, 120 and 3000 ${\rm fb}^{-1}$.
Two sigma anomaly obtained from current data is remarked by the regions closed by purple lines near $m_X=350$ GeV.
The decay width is taken to be $10^{-2}$ (solid) and $10^{-4}$ (dashed) of its mass.
In the left plot, grey dotted horizontal lines represent one stop-like particle and the equivalent of one generation and three generations of degenerate scalar quarks. In the right plot, such lines correspond to one vector-like up type quark and a degenerate set of two vector-like up and two vector-like down type quarks (motivated by one complete vector-like family).
}
\label{plotcontour}
\end{figure}

Fig.~\ref{plotcontour} shows current (dark red) and expected (red, green and blue) 95\% confidence level (C.L.) exclusion limits on $m_X$ and $C_X$ parameter space for scalar (left) and fermion (right). The integrated luminosity for expected exclusion limits are 15.4 (red), 120 (green) and 3000 ${\rm fb}^{-1}$ (blue) which represent the current, run II and high-luminosity LHC data. The solid lines correspond to $\Gamma_X/m_X=10^{-2}$ which represent the conservative limits. For larger widths, we get only slightly weaker limits \cite{Chway:2015lzg}. For smaller widths, we get significantly stronger limits from the sharper shape of the signal. In this case it is beneficial to reduce the bin size which will be possible with future data. Green and blue dashed lines indicate future sensitivity for $\Gamma_X/m_X=10^{-4}$. 
The regions inside orange contours correspond to $2\sigma$ anomaly with the best fit: $m_{S(F)}=350~(290)$ GeV and $C_{S(F)}=6.7~(2.1)$ having 2.2 (2.5) $\sigma$ significance assuming $\Gamma_X/m_X=10^{-2}$ .
In the left plot, grey dotted horizontal lines represent one stop-like particle and the equivalent of one generation and three generations of degenerate scalar quarks. In the right plot, such lines correspond to one vector-like up type quark and a degenerate set of two vector-like up and two vector-like down type quarks (motivated by one complete vector-like family).

By looking at their intersections with the blue curves, we estimate some benchmark points assuming 3 ${\rm ab}^{-1}$ of the integrated luminosity.
For $\Gamma_X/m_X=10^{-2}$, one up (two up+two down) type quark(s) lighter than 360 (460) GeV, one stop-like particle lighter than 200 GeV and the equivalent of one (three) generation(s) of supersymmetric quark partner lighter than $280$ ($390$) GeV would be probed.
For $\Gamma_X/m_X=10^{-4}$, one up (two up+two down) type quark(s) lighter than 450 (700) GeV, one stop-like particle lighter than 300 GeV and the equivalent of one (three) generation(s) of supersymmetric quark partner lighter than $480$ ($800$) GeV would be probed.
\blue{When the total decay width of the bound state is small, the integrated luminosity needed to achieve a sufficient chi square of the bin to which the resonance belongs is related with the $C_X$ limit, the total decay width and the size of the bin by $C_X \propto \left( \f{{\rm bin\, size}}{{\rm Luminosity}} \right)^{1/4} \Gamma_{\rm tot}^{1/2}$. Therefore, one stop-like particle as heavy as 300 GeV can be probed for 24 GeV bin size with the integrated luminosity of 180 ${\rm fb}^{-1}$ when the bound state dominantly decays to two gluon state so that $\Gamma_{\rm tot}=10^{-5}\times 300$ GeV. With the same parameters, the luminosity of 300 ${\rm fb}^{-1}$ is found to be required in Ref. \cite{Martin:2008sv}. We obtained smaller integrated luminosity because we assumed K-factor canceling cut selection efficiency.}

\begin{figure}[t]
\begin{subfigure}[b]{0.3\textwidth}
\includegraphics[width=1.0\textwidth]{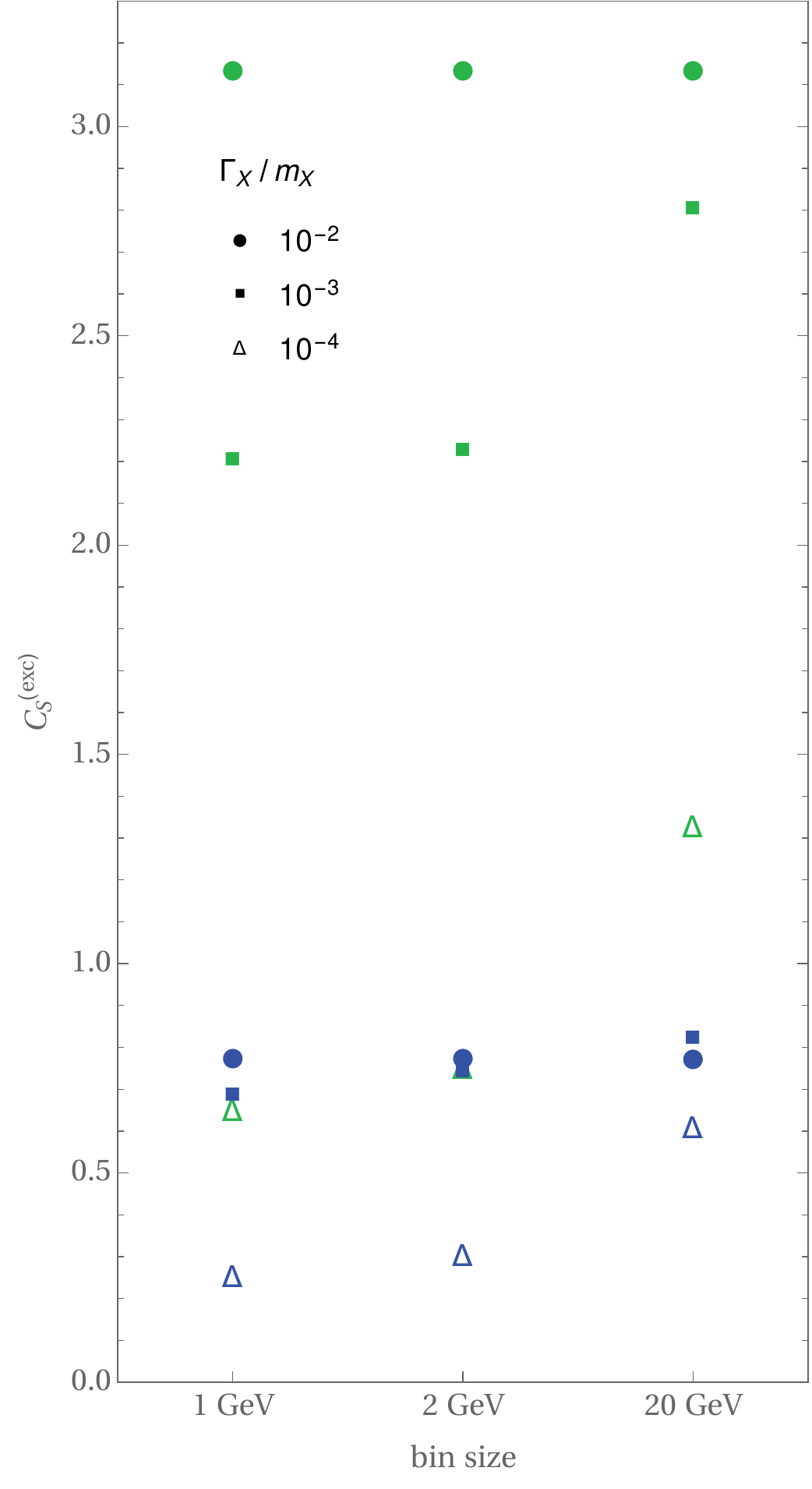}
\caption{scalar}
\label{bins}
\end{subfigure}
\begin{subfigure}[b]{0.3\textwidth}
\includegraphics[width=1.0\textwidth]{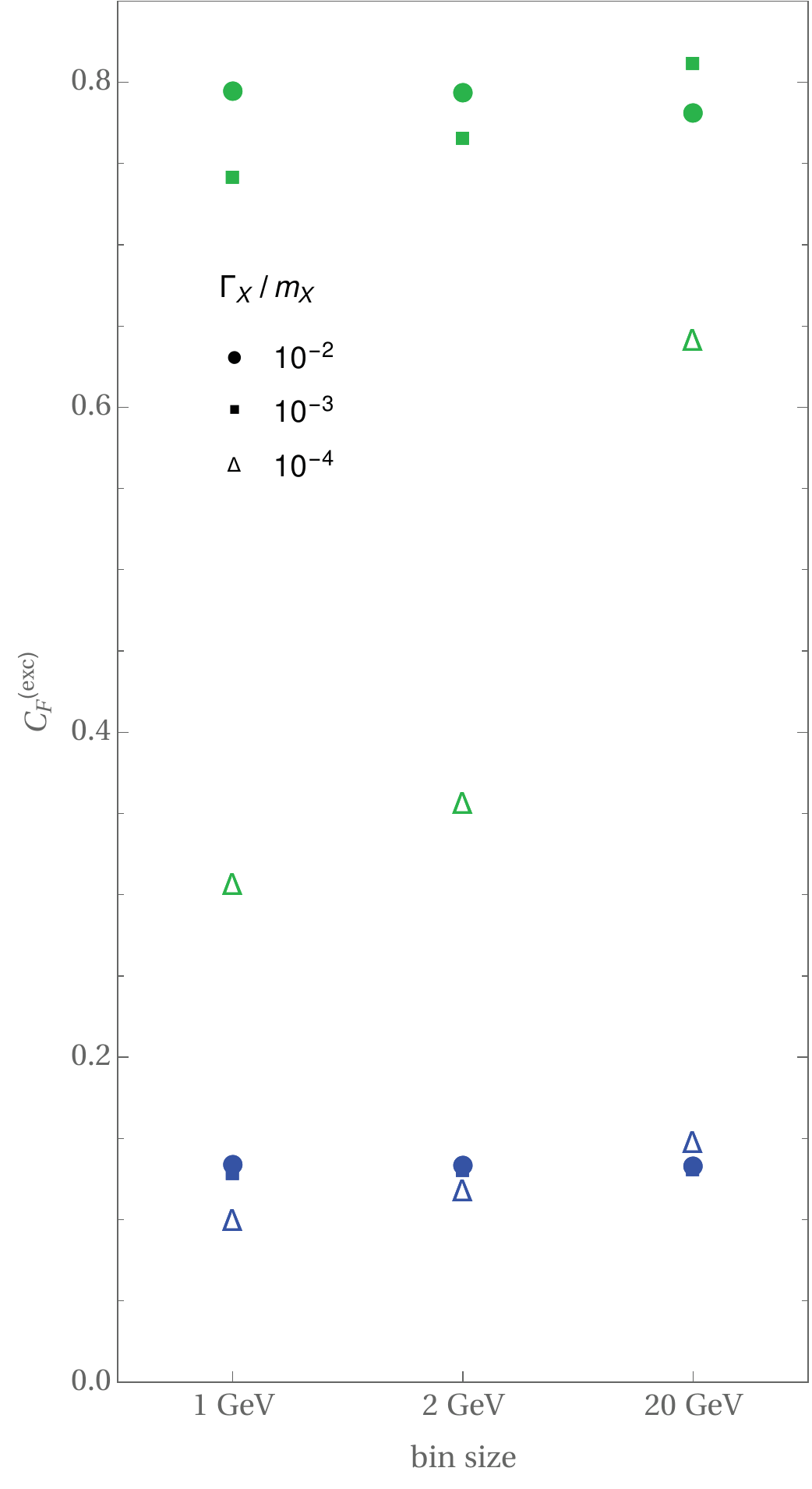}
\caption{fermion}
\label{binf}
\end{subfigure}
\caption{Expected upper bound of $C_X$ where $m_X=300$ GeV for integrated luminosity $120~{\rm fb}^{-1}$ (green) and $3~{\rm ab}^{-1}$ (blue) 
with $\Gamma_X/m_X=10^{-2}$ (circle) , $10^{-3}$ (square)
 and $10^{-4}$ (triangle).}
\label{binsize}
\end{figure}

The limits in Fig.~\ref{plotcontour} assume the bin size  20 GeV for $\Gamma_X/m_X=10^{-2}$ while for $\Gamma_X/m_X=10^{-4}$ we choose the bin size 2 GeV for 120 ${\rm fb}^{-1}$ and 1 GeV for 3 ${\rm ab}^{-1}$. 
\red{
The bin size was chosen to optimize the sensitivity.
In the previous ATLAS paper with 15 ${\rm fb}^{-1}$, they used 20 GeV bin size.
We consider that 2 GeV for 120 ${\rm fb}^{-1}$ is a reasonable choice as the integrated luminosity is about 10 times larger.
For 3000 ${\rm fb}^{-1}$, photon detector resolution is expected to be about 1 GeV.
}
In order to understand the importance of the proper choice of the bin size for different widths, we provide Fig.~\ref{binsize} in which the expected upper bound on $C_X$ is depicted assuming $m_X=300$ GeV for the integrated luminosity 120 ${\rm fb}^{-1}$ (green) and 3 ${\rm ab}^{-1}$ (blue) with $\Gamma_X/m_X=$ $10^{-2}$ (circle), $10^{-3}$ (square) and $10^{-4}$ (triangle). 
We can see that for large widths, the limits are not sensitive to the bin size.
Weak dependence on the bin size in this case indicates that the analysis relies more on the structure of loop function than bound state structure.
However, as the width decreases, the smaller bin size sets significantly stronger limits.

\red{
In the small width limit, we can compare our exclusion limit with the result obtained using the usual bound state analysis \cite{Kats:2016kuz} in which the production and the decay is separately considered. 
In this limit, the narrow width approximation should give the same result as our full resummation computation as given in Appendix \ref{appnarrow}. 
Nevertheless, 
we obtain weaker limit for $\Gamma_X \simeq 10^{-5} m_X$. There are three reasons for this. Firstly, the usual bound state analysis used leading order Coulomb potential while we use the NLO potential which produces smaller bound state amplitude as shown in Fig. \ref{Ampreal} and Fig. \ref{Ampim}. Secondly, the running of $\alpha_s$ makes the resonance peak more squeezed and we obtain smaller signal cross section compared to the narrow width approximation. Finally, the global fitting of the background shape slightly reduces the $\chi^2$ of the signal as we did not assume that we know the background precisely.}

\section{Conclusions}

In this paper, we presented detailed explanation of the threshold resummation and the leading log order matching of the one-loop result with non-relativistic effective theory. 
We showed how the diphoton invariant mass spectrum varies depending on decay width, color representation and electric charge of the new particle. 
We also included interference with the standard model quarks which is important for new particles with small combined charges. 

We presented new exclusion limits from current LHC data corresponding to 15.4 fb$^{-1}$ and projections for expected exclusion limits.
For example, assuming $\Gamma_X/m_X \le 10^{-2}$, the LHC will be sensitive to a top-like particle up to 360 GeV and a stop-like particle up to 200 GeV.
For $\Gamma_X/m_X \le 10^{-4}$,  the LHC will be more sensitive and a top-like particle up to 450 GeV and a stop-like particle up to 300 GeV can be seen. Any new particles with larger $SU(3)_{\rm C}$ representation and/or larger $U(1)_{\rm EM}$ charges can be probed in larger mass ranges.

Our exclusion limits on the combined $SU(3)_C$ and $U(1)_{\rm EM}$ charge do not depend on details of a given model
just like the limits on hypercharges of new particles from Drell-Yan process \cite{Gross:2016ioi, Goertz:2016iwa}
or limits on colored particles from the ratio of 3 to 2 jets cross section \cite{Becciolini:2014lya}.
If the new particle is colored, our projected limits are significantly stronger than those from Drell-Yan process.
In addition, if the electromagnetic charge of the new particle is not small, our limits can also exceed those from the ratio of 3 to 2 jets cross section. Furthermore, in the case the effects of a new particle are seen, our process can be used to measure the mass and the width of the new particle which is not possible using these other methods.

\noindent
{\bf Acknowledgements}
This work was supported by the National Research Foundation of Korea (NRF), No. 0426-20140009 and No. 0409-20150110. The work of RD was supported in part by the U.S. Department of Energy under grant number {DE}-SC0010120
and by  the Ministry of Science, ICT and Planning (MSIP), South Korea, through the Brain Pool Program.
The work of THJ was supported by IBS under the project code, IBS-R018-D1.

\appendix
\section{\red{Small Width Limit}}
\label{appnarrow}

\red{When the width of the new particle X is small, one can use the usual bound state analysis which gives signal cross section as the product of the production cross section and the branching ratio of the bound states. In this appendix, we show this following the appendix of Ref. \cite{Kats:2009bv}. 
}

\red{
Let a new scalar particle which is in fundamental representation of $SU(3)_C$ have electric charge $Q=1$ (i.e., $C_S=N_S T_{R_S} Q^2=1/2$). For unpolarized beam with the center of mass energy $\sqrt{S}$, the differential cross section from digluon to diphoton after the Coulomb resummation for the new particle is
\small{
\bea
\f{d\sigma_{gg\rightarrow \gamma\gamma}}{d m_{\gamma\gamma}}
=\Bigg( \int^1_{\f{m_{\gamma\gamma}^2}{S}}  f_g \left( x \right) && f_g  \left( \f{m_{\gamma\gamma}^2}{x S} \right) \f{2 m_{\gamma\gamma}}{x S} d x \Bigg) 
\f{ \alpha^2 \alpha_s^2 }{2^{12} \pi m_{\gamma\gamma}^2} \nn  \label{eq:A1}\\ 
 \times \int^1_{-1} d \cos{\theta} &&
\Bigg(
\left| \mathcal{M}_{++++}+ \mathcal{A}_{++++}-4\pi \f{4\pi}{m^2}(G-G_0) \right|^2 \nn \\
&&+\left| \mathcal{M}_{++--}+ \mathcal{A}_{++--}-4\pi \f{4\pi}{m^2}(G-G_0) \right|^2 \label{differentialcross} \\
&&+\left| \mathcal{M}_{+-+-}+ \mathcal{A}_{+-+-} \right|^2
+\left| \mathcal{M}_{+--+}+ \mathcal{A}_{+--+} \right|^2
+4\left| \mathcal{M}_{+++-}+ \mathcal{A}_{+++-} \right|^2
\Bigg),
\nn
\eea
}
where $\alpha^2 \mathcal{M}$ and $\alpha^2 \mathcal{A}$ are amplitudes of diphoton to diphoton mediated by quarks and the new scalar, respectively. (If the new particle is a fermion, the Green's function terms should be doubled.) Defining the glue-glue parton luminosity as
\bea
\mathcal{L}_{gg}\left( m_{\gamma\gamma}^2 \right) = \f{m_{\gamma\gamma}^2}{S} \int^1_{m_{\gamma\gamma}^2/S}
\f{d x}{x} f_{g/p} \left( x \right) f_{g/p} \left( \f{m_{\gamma\gamma}^2}{x S} \right),
\eea
the integral in the first big parenthesis of Eq. \eqref{eq:A1} is $ 2 \mathcal{L}_{gg}\left( m_{\gamma\gamma}^2 \right) / m_{\gamma\gamma}$.
}
\red{When the decay width of the particle is small, the Green's function of the bound state resonances can be approximated as
{\small\bea
&&  G(E+i\Gamma,\vec{x}=\vec{0})     \\
&&  = - \frac{m_X^2}{4\pi} \left\{ \sqrt{-\frac{E+i\Gamma}{m_X}} 
 -C_C \bar{\alpha_s}(\mu) \ln \left(\mu\sqrt{\frac{1}{-m_X (E+i\Gamma)}}\right)
  -\frac{2}{\sqrt{m_X}}\sum_{n=1}^{\infty}\frac{E_n}{\sqrt{(-E-i \Gamma)}-\text{sign}(C_C)\sqrt{E_n} }\right\}  \nn \\
  &&\simeq - \frac{m_X^2}{4\pi} \left\{
  \frac{2}{\sqrt{m_X}}\sum_{n=1}^{\infty}\frac{2 E_n^{3/2}}{E+E_n+i \Gamma }\right\} \nn \\
  &&\simeq
  - \sum_{n=1}^{\infty}\frac{ 2M_n \f{C_C^3 \bar{\alpha_s}^3 m_X^{3}}{8\pi n^3}}{m_{\gamma\gamma}^2-M_n^2+2i M_n \Gamma } ,
\nn 
\eea}where $E_n=C_C^2 \bar{\alpha_s}^2 m_X/4n^2$, $C_C=4/3$, $m_{\gamma\gamma}^2=(2m_X+E)^2$, and $M_n=2m_X-E_n$. Since the peak height decreases rapidly as $1/n^3$, taking only the $n=1$ term, we find
\footnote{One caution is that if the scale $\mu$ of $\bar{\alpha}_s(\mu)$ runs as a function of $m_{\gamma\gamma}$,  $M_1(\bar{\alpha}_s(\mu) )$ is no longer a constant and the approximate equality to Eq. \eqref{A5} does not hold as the function is no more a Lorentzian function. If one uses NLO potential, the scale dependence is reduced and the error of the approximation as a Lorentzian delta function for running $\mu$ becomes smaller. }
\bea
  \left| G(E+i\Gamma,\vec{x}=\vec{0}) \right|^2   
  &&\simeq
  \left|- \frac{ 2M_1 \f{C_C^3 \bar{\alpha_s}^3 m_X^{3}}{8\pi }}{m_{\gamma\gamma}^2-M_1^2+2i M_1 \Gamma } \right|^2 \\
  &&\simeq \left(\f{C_C^3 \bar{\alpha_s}^3 m_X^{3}}{8\pi}\right)^2 \f{M_1}{\Gamma} 2\pi \delta \left( m_{\gamma\gamma}^2-M_1^2 \right)  \label{A5}
  .
\eea
With this approximation, the signal is given by
\bea
\int d\sigma_{gg\rightarrow \gamma\gamma}
&&\simeq \int d m_{\gamma\gamma}^2 \f{ \mathcal{L}_{gg}\left( m_{\gamma\gamma}^2 \right)}{m_{\gamma\gamma}^2}
\f{ \alpha^2 \alpha_s^2 }{2^{12}  \pi m_{\gamma\gamma}^2}  4
\left(
 \f{(4\pi)^2}{m_X^2} 
\right)^2 
\left(\f{C_C^3 \bar{\alpha_s}^3 m_X^{3}}{8\pi }\right)^2 \f{M_1}{\Gamma} 2\pi \delta \left( m_{\gamma\gamma}^2-M_1^2 \right) \nn \\
&&= \f{ \mathcal{L}_{gg}\left( \left(2m_X\right)^2 \right)}{ \left(2m_X\right)^2}
\f{ \alpha^2 \alpha_s^2 }{2^{12}  \pi \left(2m_X\right)^2}  4
\left(
 \f{(4\pi)^2}{m_X^2} 
\right)^2 
\left(\f{C_C^3 \bar{\alpha_s}^3 m_X^{3}}{8\pi }\right)^2 \f{M_1}{\Gamma} 2\pi 
. 
\label{differentialcrosssmall}
\eea
Remind that $\Gamma$ is the decay width of the scalar particle itself. In terms of the total decay width, $\Gamma_{tot}$, we should substitute $\Gamma_{tot}/2$ for $\Gamma$ as in the appendix of Ref. \cite{Kats:2009bv}. Plus, considering the annihilation rate gives $\Gamma_{tot}=2\Gamma+\Gamma_{ann}$. In order to compare the result with Refs. \cite{Kats:2012ym, Kats:2016kuz}, we set $\Gamma=0$. From Refs. \cite{Martin:2008sv, Kats:2012ym}, the annihilation decay width is
\bea \Gamma_{ann}(\eta\rightarrow 2g)&&= \f{4\pi}{3} \left( \f{\alpha_s}{m_X} \right)^2 \left| \psi (0) \right|^2.  \\ 
&&=\f{1}{6} \alpha_s^2 C_C^3 \bar{\alpha_s}^3 m_X .
 \eea
(If the constituent particle is a fermion, this should be doubled.)
Substituting $\Gamma_{ann}/2$ for $\Gamma$ results in
\bea
\int d\sigma_{gg \rightarrow \gamma\gamma}
&&\simeq \f{ \mathcal{L}_{gg}\left( \left(2m_X\right)^2 \right)}{ \left(2m_X\right)^2}
\f{\alpha^2 \alpha_s^2 }{2^{12} \pi \left(2m_X\right)^2}  4
\left(
 \f{(4\pi)^2}{m_X^2} 
\right)^2 
\left(\f{C_C^3 \bar{\alpha_s}^3 m_X^{3}}{8\pi }\right)^2 \f{M_1}{\f{1}{12} \alpha_s^2 C_C^3 \bar{\alpha_s}^3 m_X} 2\pi  \nn
\\
&&=\f{3 C_C^3}{2^{8}}\pi^2 \alpha^2 \bar{\alpha_s}^3\f{\mathcal{L}_{gg}}{m_X^2}
. 
\eea
Recovering electric charge $Q$, we finally obtain
\bea
\int d\sigma_{gg \rightarrow \gamma\gamma}
\simeq \f{3 Q^4 C_C^3}{64}\pi^2 \alpha^2 \bar{\alpha_s}^3\f{\mathcal{L}_{gg}\left( \left(2m_X\right)^2 \right)}{(2m_X)^2}
.
\eea
This agrees with Eq. (4.3) in Ref. \cite{Kats:2012ym} and Eq. (2.6) in Ref. \cite{Martin:2008sv}.
}

\section{\red{Procedure for Expected Exclusion Limits}}
\label{LLapen}

Here, we describe in detail how we obtained the expected exclusion curves:

\begin{enumerate}
\item{Non-gluon initiated process was estimated from the ATLAS fitted plot.}

\red{We read the differential cross section of $pp\rightarrow\gamma\gamma$ from the background only fit in the figure 4 of the ATLAS note \cite{ATLAS:2016eeo}. After subtracting the differential cross section of $gg\rightarrow\gamma\gamma$ from it, we fitted (in log scale as the figure is in log scale) non-gluon initiated process as ${N_0} (1-x^{1/3})^{b_0} x^{{a_{00}}}$ resulting in $a_{00} =-3.67$ and $b_0=4.15$.}

\item{Gluon initiated process for non-zero $C_X$ was determined.}

\red{Perturbatively, we can trust the one loop differential cross section for energy range that gives the velocity $\beta$ of the particle X with $\f{C_C \alpha_s}{\beta}<\f{1}{4}$. On the other hand, we can trust the Coulomb resummed result within the range $\beta<\f{1}{4}$ as it was calculated in the non-relativistic limit. For regions which do not belong to any of the two, we have to interpolate the one loop result and resummed result. In the future if both continuum calculation and resummed calculation are done in higher order, this arbitrariness will be lessened. For now, we have to choose among various interpolation choices such as using, in the interpolation region, the one loop result, the resummed result, or a linear interpolation of the two. As an alternative, we can choose to use resummed result in the interpolation region and then shift the one loop result horizontally to make the differential cross section continuous. We worked in this interpolation because discontinuities can give rise to artificial shapes affecting likelihood estimation and overall shift can be compensated by the fitting that we use to estimate non-gluon initiated process in the next step. }

\item{The exclusion limit was obtained by maximum binned likelihood estimation.}

\red{We binned the differential cross section as 
\small{
\bea
\mu_i (C_X,a_0,b) = {\rm Lum} \int_{\rm i-th \,\, bin} dm_{\gamma\gamma} \left[ {N} \left(1-\left(\f{m_{\gamma\gamma}}{\sqrt{S}}\right)^{1/3}\right)^{b} \left(\f{m_{\gamma\gamma}}{\sqrt{S}}\right)^{{a_{0}}} + \f{d \sigma (gg\rightarrow\gamma\gamma)}{dm_{\gamma\gamma}}\right],
\eea}
where $\rm Lum$ is the integrated luminosity and the $gg\rightarrow\gamma\gamma$ differential cross section is as determined in the previous step. Then, 95\% C.L. expected exclusion limits were obtained by
\bea
(1.96)^2 = -2 \ln  \f{L(C_X,\hat{a_0},\hat{b})}{L(C_X=0,a_{00},b_0)},
\eea
where the likelihood $L(C_X,a_0,b)=\Pi_i P( \mu_i(0,a_{00},b_0) | \mu=\mu_i(C_X,a_0,b) )$ is the products of Poisson probability to find $\mu_i(0,a_{00},b_0)$ with a mean value $\mu_i(C_X,a_0,b)$. The $\hat{a_0}$ and $\hat{b}$ maximize the likelihood for a given $C_X$. The normalization of the fitting function is always chosen to keep the total number of events to remain the same. In the future, we believe that generating the non-gluon initiated process like in Ref. \cite{Campbell:2016yrh} without relying on fitting function will be possible and more strict exclusion limit will be obtained.}

\end{enumerate}

\end{document}